\documentclass[journal]{IEEEtran}
\usepackage{amsmath,amssymb,physics,tabularx,graphicx,multirow,hhline,algorithm,mathtools,algpseudocode,varwidth,color,soul,amsthm,cite,bbm,mwe}
\usepackage{nomencl}
\makenomenclature
\usepackage[skins,breakable,hooks,theorems]{tcolorbox}
\PassOptionsToPackage{hyphens}{url}
\usepackage{hyperref}

\algnewcommand{\algorithmicforeach}{\textbf{for each}}
\algdef{SE}[FOR]{ForEach}{EndForEach}[1]
  {\algorithmicforeach\ #1\ \algorithmicdo}
  {\algorithmicend\ \algorithmicforeach}
\newsavebox{\ieeealgbox}

\ifCLASSINFOpdf
\else
\fi

\tcbset{
  highlight math style={
    colback=yellow,
    arc=0pt,
    outer arc=0pt,
    boxrule=0pt,
    top=2pt,
    bottom=2pt,
    left=2pt,
    right=2pt,
  }
}

\ifCLASSOPTIONcompsoc \usepackage[caption=false,font=normalsize,labelfont=sf,textfont=sf]{subfig}
\else
\usepackage[caption=false,font=footnotesize]{subfig}
\fi
\usepackage{float}

\begin{document}
\algnewcommand{\algorithmicgoto}{\textbf{go to}}%
\algnewcommand{\Goto}[1]{\algorithmicgoto~\ref{#1}}%
\title{On \textcolor{black}{an Information and} Control 
Architecture for Future Electric Energy Systems}

\author{Le Xie,~\IEEEmembership{Fellow,~IEEE},
        Tong Huang, ~\IEEEmembership{Member,~IEEE},
        P. R. Kumar, ~\IEEEmembership{Life Fellow,~IEEE},
        Anupam A. Thatte, ~\IEEEmembership{Member,~IEEE},
        and Sanjoy K. Mitter, ~\IEEEmembership{Life Fellow,~IEEE}

\thanks{L. Xie and P. R. Kumar are with the Department of
Electrical and Computer Engineering, Texas A\&M University, College Station, TX, USA. Email: \text{le.xie@tamu.edu}.  This work was supported in part by the U.S. Department of Energy’s Office of Energy Efficiency and Renewable Energy (EERE) through the Solar Energy Technologies Office (SETO) under Grant DE-EE0009031, and in part by the National Science Foundation under Grant OAC-1934675 and ECCS-2035688.}
\thanks{T. Huang is with the Department of Electrical and Computer Engineering, San Diego State University, San Diego, CA, USA.}
\thanks{A. A. Thatte is with the Midcontinent Independent System Operator (MISO), Carmel, IN, USA. Views expressed in this article are solely of the author and do not necessarily reflect those of the Midcontinent Independent System Operator.}
\thanks{S. K. Mitter is with the Laboratory for Information and Decision Systems, Massachusetts Institute of Technology, Cambridge, MA, USA.}
}


\maketitle

\begin{abstract}
This paper presents considerations towards \textcolor{black}{an information and} control architecture for future electric energy systems driven by massive changes resulting from the societal goals of decarbonization and electrification. 
\textcolor{black}{This paper} describes the
 new requirements and challenges 
 of an extended 
 \textcolor{black}{information} and control architecture that need to be addressed for continued reliable delivery of electricity. 
It identifies several new \textcolor{black}{actionable information }and control loops, along with their spatial and temporal scales of operation, which can together meet the needs of future grids and enable deep decarbonization of the electricity sector.
The present architecture of electric power grids designed in a different era is thereby extensible to allow the incorporation of increased renewables \textcolor{black}{and other emerging electric loads}. 

\end{abstract}

\nomenclature{TCP/IP}{Transmission Control Protocol/Internet Protocol }
\nomenclature{ISO}{Independent System Operator}
\nomenclature{GENCO}{Generation Company}
\nomenclature{VPP}{Virtual Power Plant}
\nomenclature{DISCO}{Distribution Company}
\nomenclature{TRANCO}{Transmission Company}
\nomenclature{EV}{Electric Vehicle}
\nomenclature{EVCS}{Electric Vehicle Charging Station}
\nomenclature{GRIP}{Grid with Intelligent Periphery}
\nomenclature{RTO}{Regional Transmission Organization}
\nomenclature{PUC}{Public Utility Commission}
\nomenclature{SCADA}{Supervisory Control and Data Acquisition}
\nomenclature{AVR}{Automatic Voltage Regulator}
\nomenclature{AGC}{Automatic Generation Control}
\nomenclature{ED}{Economic Dispatch}
\nomenclature{UC}{Unit Commitment}
\nomenclature{FERC}{Federal Energy Regulatory Commission}
\nomenclature{CAISO}{California Independent System Operator}
\nomenclature{NYISO}{New York Independent System Operator}
\nomenclature{DER}{Distributed Energy Resources}
\nomenclature{dLMP}{\textcolor{black}{Distribution} Locational Marginal Prices}
\nomenclature{PHS}{Pumped Hydroelectric Storage}
\nomenclature{IBR}{Inverter-Based Resource}
\nomenclature{PMU}{Phasor Measurement Unit}
\nomenclature{FCI}{Fault Circuit Indicator}
\nomenclature{DC}{Direct Current}
\nomenclature{AC}{Alternating Current}
\nomenclature{DG}{Distributed Generation}
\nomenclature{PCC}{Point of Common Coupling}
\nomenclature{OPF}{Optimal Power Flow}
\nomenclature{PLL}{Phase Lock Loop}
\nomenclature{PWM}{Pulse Width Modulation}
\nomenclature{BES}{Battery Energy Storage}
\nomenclature{IoTs}{Internet of Things}
\nomenclature{$\mu$MS}{Microgrid Management System}
\nomenclature{SOC}{State of Charge}
\nomenclature{V2G}{Vehicle to Grid}
\nomenclature{V2H}{Vehicle to Home}
\nomenclature{LSE}{Load Serving Entity}
\nomenclature{MES}{Multi-energy Systems}
\nomenclature{CHP}{Combined Heat and Power}
\nomenclature{DR}{Demand Response}
\nomenclature{ANN}{Artificial Neural Network}
\nomenclature{VSI}{Voltage Source Inverter}
\nomenclature{OLTC}{On-load Tap Changer}
\nomenclature{TOU}{Time of Use}
\nomenclature{PID}{Proportional–Integral–Derivative}

\section{Introduction} \label{Section1}



Electric grids around the world are undergoing massive changes. At the heart of these changes is the need to decarbonize the electricity sector by replacing fossil fuels with renewables or other carbon-free energy resources. {The latest report from the United Nations Intergovernmental Panel on Climate Change (IPCC) shows the threats from increasing global temperature and the urgency to mitigate greenhouse gas emissions \cite{Portner2022climate}.} Simultaneously, the same goal is driving the vehicular sector towards increasing electrification, compounding the challenge placed on the electricity sector. This worldwide goal of decarbonization is driving several changes. Unlike fossil fuels, solar/wind power generation cannot be increased when needed. Hence power consumption may need to be reduced on occasion when solar/wind power is insufficient to meet demand. How such reduction can be induced on the part of energy consumers poses challenges.
One possibility is through pricing, which itself raises, on the one hand, issues concerning elasticity of consumers, informing consumers of real-time prices, taking account of human behavior, and, on the other hand, of sensing what devices are on/off, consuming how much power, how they can be controlled, {\color{black}{and}} the loss of privacy when individual consumer information is communicated elsewhere.  It also entails a greater role for \textcolor{black}{energy} storage, to provide a buffer between generation and consumption. Such storage can be \textcolor{black}{provided by} batteries, by pumping water to higher levels, or even in the form of heat in buildings. At the same time, technologies such as power electronics offer many additional possibilities such as grid-forming capabilities that would allow for \textcolor{black}{the} provision of essential grid support services for frequency and voltage, despite much lower rotating mass in the power system as coal-fired synchronous generation is being retired in a number of countries. 

Controlling all these complex next-generation interactions requires an examination of the important issue of the {\textit{architecture of information and control loops}} of future power systems \cite{taft2016grid}. {The success of the present power grid is due to several well-conceived information and control loops that have played a critical role in the reliable delivery of power. These include
Automatic Voltage Regulators, droop control, Automatic Generation Control, Economic Dispatch, and Unit Commitment.  
Can the present \textcolor{black}{information and} control architecture be extended to support the oncoming changes? If so, what additional control loops are necessary, and at what spatial and temporal scales? Addressing this central architectural challenge is the goal of this paper. It suggests several specific control loops with the addition of which the present \textcolor{black}{information and} control architecture will be extensible to support the reliable delivery of power in the future grid. For each control loop this paper restricts itself to describing its inputs, outputs, and objective or purpose, but refrains from mentioning any specific control methodologies to be used for design. There are usually several alternative approaches that may be employed, and the optimal design of such control laws is left for future research.}

{Elaborating on the above, i}n power systems it has been its 
{\textit{\textcolor{black}{information and} control architecture}} that has provided reliable delivery of power in spite of many
uncertainties and disruptions at several levels.
Power systems are enormously complex. They involve phenomena spanning time scales ranging from microseconds (e.g., for power electronics switching), to milliseconds (e.g., for electromagnetic phenomena), to seconds (e.g., for electromechanical phenomena governing spinning generators), to minutes (e.g., for dispatch of generators), to hours (e.g., for startup/shutdown of generators), to years (e.g., for building of new transmission lines). Disruptions happen at all times scales, {e.g.}, lightning strikes, failure of equipment, changes in customer loads, changes in weather that affect demand, diurnal and seasonal changes in both demand and renewables, {and} gas pipeline disruptions. Similarly, there is constant change in phenomena spanning an extremely wide spatial scale, from individual households in small neighborhoods interconnected by a distribution feeder line to massive power plants connected by transmission lines traversing hundreds of miles. Nevertheless power systems have provided a reliable electricity supply, i.e., the desired amount of power that each user needs and with associated variables (such as voltage, frequency, power factor, harmonic distortion) that define power quality being within tolerance limits from their nominal values. This success is fundamentally due to the multiple control loops forming the 
\textcolor{black}{information and} control architecture that regulate all the electrical variables in spite of all the uncertainties and disturbances. 



The basis underpinning the \textcolor{black}{information and} control architecture in power systems is the separation of temporal and spatial scales. This separation has clarified the tasks that control loops at several time and spatial scales have to support. Successful design methodologies at each time-scale and spatial scale have been designed, which, when composed together, have resulted in a very reliable overall system.  The control loops have also been so designed that there are no unintended adverse interactions between different control loops. Moreover, importantly, the responsibilities for the different control loops have been appropriately assigned and aligned with the economic arrangements of all the entities involved, \textcolor{black}{for example, generation, transmission and distribution companies, and consumers}. This has led to an efficient and reliable electric energy system.  This existing architecture is briefly described in Section \ref{Sec:OldGrid}. 

The resulting design methodologies have allowed the physical system to grow in size.
Indeed, it is the motivation to meet the high standard set by power systems that has led to the galvanizing slogan ``plug and play" used in other technological domains. Moreover, the architecture has also allowed the evolution of improved control loop design methodologies over time \cite{xie2021Joule}. The evolution of voltage control from regional-based to inter-regional coordinated is a very good example of how control loop design methodology has changed over time under the \textcolor{black}{information and} control architecture of spatial and temporal separation \cite{ilic1988secondary,sun2019review}. This development of voltage control across large-scale power systems bear similarities with the development of Transmission Control Protocol/Internet Protocol (TCP/IP) for the Internet.

\textcolor{black}{As mentioned above, power systems are undergoing a great transition with} the shift from fossil fuels to renewable energy sources such as solar and wind power, besides more traditional renewable sources such as hydro. Unlike fossil fuel power plants, as demand increases, wind/solar power cannot be increased as needed -- it is not ``dispatchable". Thus the balance between generation and consumption will have to be maintained by adjusting consumption \textcolor{black}{or utilizing storage}. One way to try to adjust consumption could be through pricing. That raises many issues such as the uncertainty in how humans react to price changes, assuming that they are even exposed to and aware of price changes. Other possibilities could be direct control of consumer consumption through, say, controlling their air conditioners/heaters, but that requires sensing of customer side variables, which involves issues such as privacy, as well as novel contracts with consumers. Buildings account for a large portion of energy consumption, and shaping their consumption to match renewable generation is essential. Buildings with their thermal inertia, batteries, including those in electric vehicles, and pumped hydro, all provide storage that can be used at appropriate time-scales to buffer generation and consumption. Simultaneously, the massive transformation in transportation electrification \textcolor{black}{will impose huge loads as well as variability on distribution system transformers}. Also, as individual homes are equipped with solar roof panels, power generation will become distributed as well as highly variable. How to maintain a reliable electricity supply come rain or shine is a major challenge. 

On the other hand, there are new technologies such as phasor measurement units (PMUs) that allow measurement of phase with high precision providing precise situational awareness of power flows.  There is also the still untapped potential of power electronics devices in providing control at a finer temporal granularity. {\color{black}{Further, the}} increasing deployment of smart meters at customers' homes will allow more complex pricing, e.g., time-of-use \cite{borenstein2002dynamic} or EnergyCoupon \cite{ming2020prediction}, as well as allow privacy constrained data gathering in real-time for more intelligent control. However, with these new technologies comes the challenge of security of the massive cyber-physical system, as information from sensors can be distorted by malicious agents.

How can the aforesaid changes be accommodated in future electric energy systems? That is the challenge addressed in this paper. There are certain requirements that should guide the evolution of the \textcolor{black}{information and} control architecture of the electric grid. First, the changes made to the existing \textcolor{black}{information and} control architecture to accommodate increasing renewables must continue to deliver electric power \textcolor{black}{both} reliably and efficiently, in spite of increased variability. Second, the \textcolor{black}{information and} control architecture must be well aligned with all the old as well as the new business entities that are emerging, such as aggregators. Third, any changes will need to be backward compatible in view of the tremendous existing investment in infrastructure. Last, but not least, the design of the architecture should continue to allow room for innovation and not stifle it. With these goals in mind, some key additions to the existing \textcolor{black}{information and} control architecture are discussed in Sections V and VI. Where possible, it is desirable to allow distributed decision making in order to foster creativity ``at the edge," in Internet parlance. However, in contrast, in some cases, such as \textcolor{black}{in} gas-grid and electric-grid operations, greater coordination is desirable to increase system reliability. 

\begin{figure*}
    \centering
    \includegraphics[width = 5in]{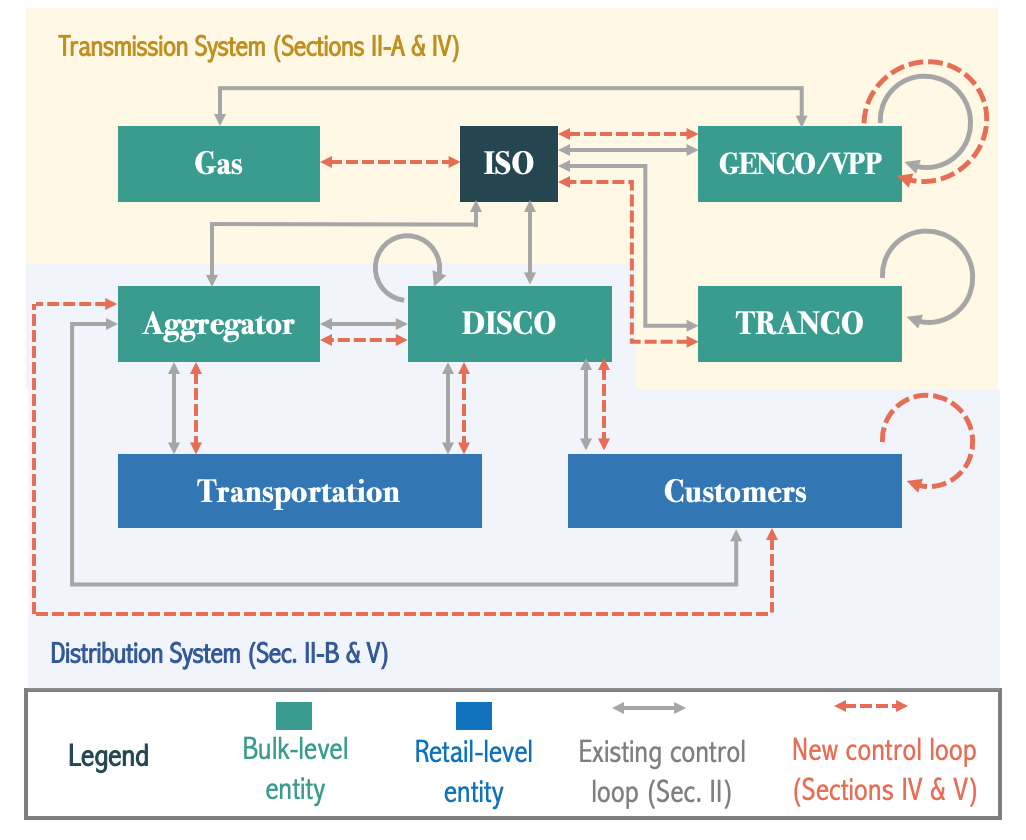}
    \caption{\textcolor{black}{Extensibility of architecture of the electric grid: The present and possible additional future control loops for incorporation of increased renewables and electric vehicles}}
    \label{fig:overview}
\end{figure*}

This paper presents the grid architecture \textcolor{black}{comprised of the} collection of control loops among key entities in the power grids. As is shown in Figure \ref{fig:overview}, the entities in the grid include {\color{black}{Independent System Operators}} (ISOs), Gas/Fuel suppliers, {\color{black}{Generation companies}} (GENCOs) or {\color{black}{Virtual power plants}} (VPPs), aggregators, {\color{black}{Distribution companies}} (DISCOs), {\color{black}{Transmission companies}} (TRANCOs), transportation components, i.e., {\color{black}{Electric Vehicles}} (EVs) and {\color{black}{Electric Vehicle charging stations}} (EVCSs), and electricity end users, i.e., customers. The interactions among different entities are identified in Figure \ref{fig:overview} where the existing control loops are indicated by solid arrows, while the new control loops to incorporate increased renewables are indicated by dashed arrows. The latter represent  research opportunities for the future grid, which are elaborated on in Section IV.

There are several references that discuss the design of power grid architectures \cite{liacco1967adaptive,1519722,WU2015436,6102397,LIU2020778,national2021future,overbye2021}. 
Dy Liacco \cite{liacco1967adaptive} introduced a state-transition viewpoint of power transmission system operation which underpins the Energy Management System of modern bulk power grids. Wu \emph{et al.} \cite{1519722} reviews the evolution of control centers in modern 
power systems.
References \cite{WU2015436,6102397} and \cite{LIU2020778} introduce a Grid with Intelligent Periphery (GRIP) architecture where power grid operation is decomposed into hierarchically structured clusters that include transmission grids, distribution grids, microgrids, and smart homes. 
Reference \cite{national2021future} identifies three dimensions that can assess and describe the grid evolution given the drivers of change in electric energy systems. Reference \cite{overbye2021} reviews the past and present power grids, and points out key technologies that are needed to be developed for supporting the operation of future transmission systems. \textcolor{black}{Reference \cite{hatziargyriou2020electricity} provides a comprehensive review for electricity infrastructure.} As the existing efforts on grid architecture design suggest, there is still a need for a holistic architecture that reflects the roles of key stakeholders in both power transmission and distribution systems \textcolor{black}{as well as the associated research needs in a concise manner.}

The remainder of this paper is organized as follows. \textcolor{black}{Section \ref{Sec:OldGrid} describes} the present \textcolor{black}{information and} control architecture, as summarized in Figure \ref{fig:overview}. Section \ref{Sec:change} elaborates the drivers of change that are happening in the electric energy systems. Section \ref{Sec:Challenges} summarizes the key research challenges on designing an \textcolor{black}{information and} control architecture for the evolving power grid with increased renewables. Sections \ref{Sec:NewLoopsTran} and \ref{Sec:NewLoopDist} describe the new control loops in tranmsission and distribution systems, respectively.  Section \ref{Sec:Conclusion} provides some concluding remarks. \textcolor{black}{The list of key acronyms is provided in the Appendix.}

\section{The Present Architecture of Electric Energy Systems}\label{Sec:OldGrid}



Before embarking on {\color{black}{discussing
the future challenges, this section 
describes the present architecture of power systems and the entities involved. }}
Given the complexity of the system, there are multiple perspectives of the architecture. One elucidates the entities involved in transmission and distribution, and the planning and operational aspects involved. 
Another perspective is through the lens of temporal and spatial scales. 
Finally, and most important functionally, is to understand the set of control loops that achieve reliable and efficient power delivery,
and how they are organized.
{\color{black}{The following subsection describes the entities involved.}}

\subsection{Entities in Transmission/Distribution Grids} \label{sec:entities}

\begin{itemize}
    \item Generation  companies (GENCOs) or virtual power plants (VPPs): A GENCO is a company that owns/operates an electric power plant. A VPP is a collection of decentralized power resources (including generation units, storage or demand response) that are controlled via software and can mimic a traditional power plant. 
    \item Distribution Companies (DISCOs): These are responsible for the supply of electricity to consumers.
    \item Transmission Companies (TRANCOs): These are the entities that own/operate the high-voltage transmission grid.
    \item Independent System Operators (ISOs): These are independent organizations that coordinate the generation and transmission of electricity across large geographic areas.
    \item Fuel (gas) Suppliers: This is the group of entities from which the fossil fuel generators purchase their fuel. For the case of natural gas this could be a producer, a local distribution company or a marketer.
    \item Customers (retail): These are the end users of electricity at the distribution level.
    \item Aggregators: An electric aggregator acts  as an agent on behalf of a collection  of  customers\footnote{Aggregators may not possess physical assets for generating and storing electric energy, whereas the VPPs typically own physical assets.}, for the purpose of negotiating the  rate with an electric supplier. 
    \item Transportation: This category covers components of the transportation infrastructure that interact with the electric grid, and include EVs and EVCSs.
\end{itemize}

\subsection{The Temporal and Spatial Scales}
\label{sec:scales}
\begin{figure}
    \centering
    \includegraphics[width = 0.6\linewidth]{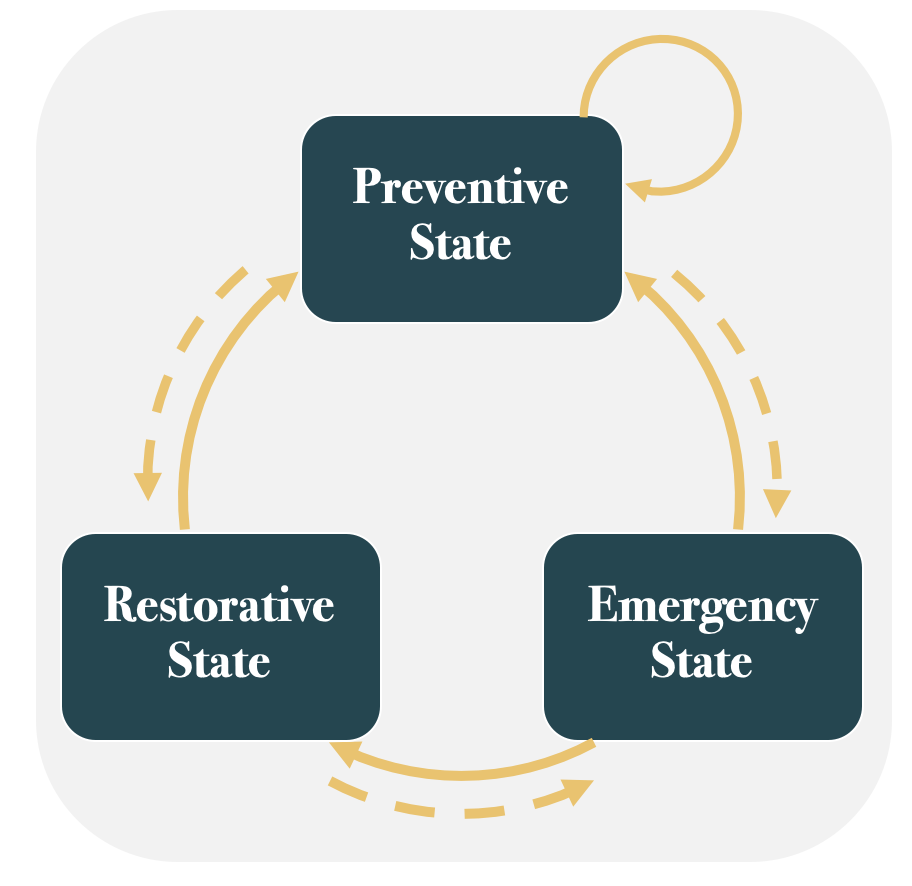}
    \caption{The Dy Liacco control center architecture}
    \label{fig:DyLiacco}
\end{figure}

\begin{figure}
    \centering
    \includegraphics[width = \linewidth]{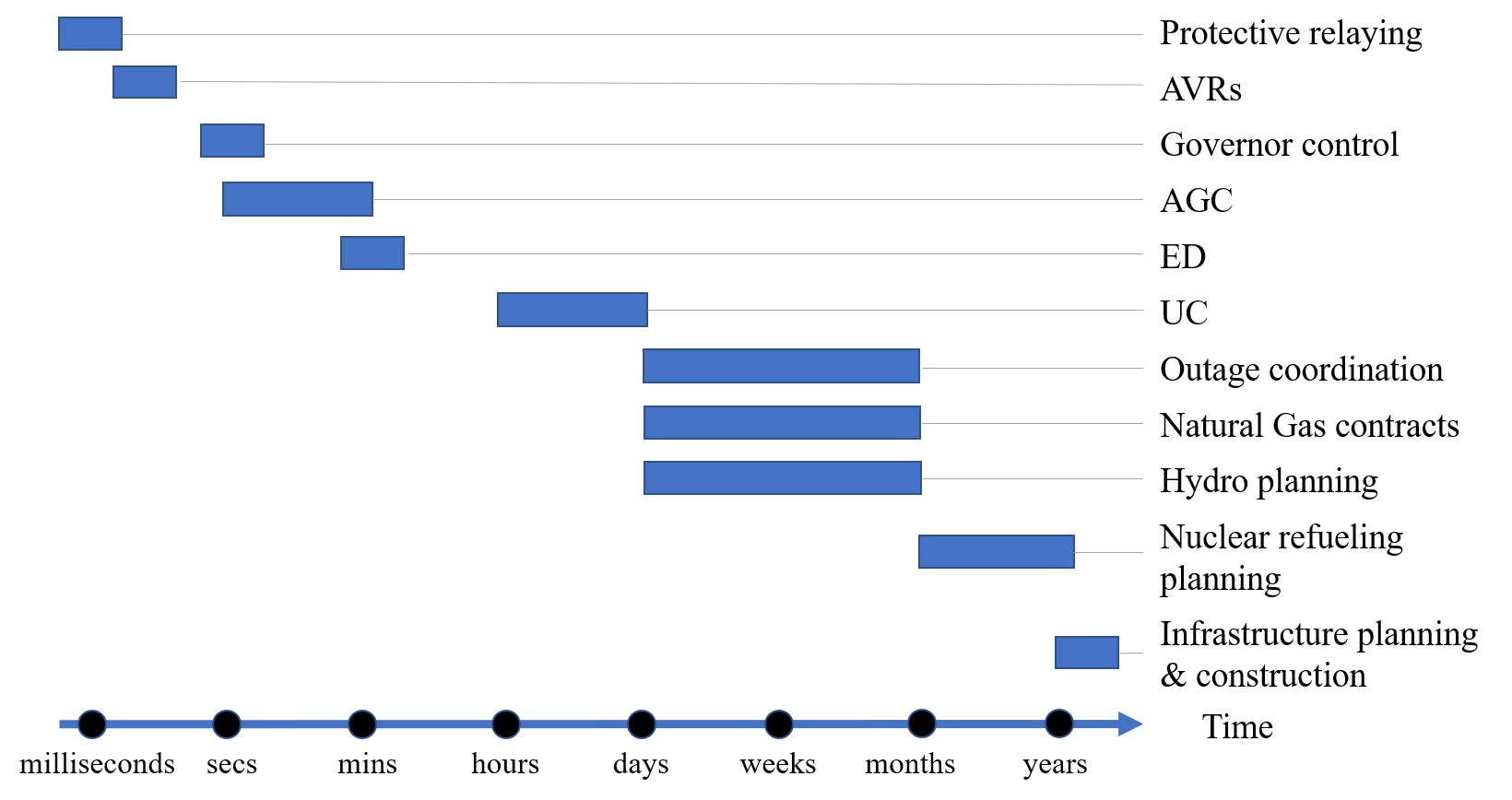}
    \caption{The timescale separation of control loops of bulk power systems}
    \label{fig:timescales}
\end{figure}

Dy Liacco \cite{liacco1967adaptive} pioneered the state transition diagram of control areas, as illustrated in Figure \ref{fig:DyLiacco}. This, together with the development of singular perturbation techniques \cite{chow1990singular}, provided the architectural and theoretical justification of spatial and temporal scale separation (shown in Figure \ref{fig:timescales}). It led to the development of a suite of decision-support tools such as steady-state analysis, small-signal dynamic security assessment, and large-signal transient stability assessment. 

The physics of the electric grid combined with the controls results in a wide ranging set of dynamics across multiple temporal scales, which are as follows:
\begin{itemize}
    \item At the fastest time-scale, of the order of microseconds, {\color{black}{is}} power electronics switching.
    \item Then {\color{black}{comes}} the order of milliseconds, for the electro-magnetic transient dynamics. 
    \item Next {\color{black}{is the scale of}} electro-mechanical dynamics, of the order of seconds, typically  associated with large rotating machines like synchronous generators. 
    \item Then the scale of power balancing 
    between generation and consumption which ranges from minutes to days, and which spans frequency regulation, dispatch and commitment actions.
    \item Power system restoration is the process of returning grid equipment to normal service following an outage. Depending on the severity of the damage, restoration time can take hours or days, or even longer in some cases of severe damage to the infrastructure.
    \item Finally at the slowest time-scale of years or decades is where {\color{black}{infrastructure planning and construction occurs}}.
\end{itemize}


Traditionally these problems have been treated separately under an assumption that the control loops in the faster time scale will be sufficient to achieve steady-state from the perspective of the slower time-scale, thereby simplifying the problem. 

Likewise, there has also been a separation of spatial scales, as follows: 1) The regional, i.e., transmission systems;
and 2) The local, i.e., distribution systems.
These spatial scales are operated somewhat separately although they are physically coupled. For example, when making decisions about the delivery of bulk electricity from generating stations to load centers, the local voltage control issues are not directly considered. Hence, for instance, the market mechanisms at the transmission level can be modified without considering the specifics of the distribution system devices.

\begin{figure*}
    \centering
    \includegraphics[width = 0.8\linewidth]{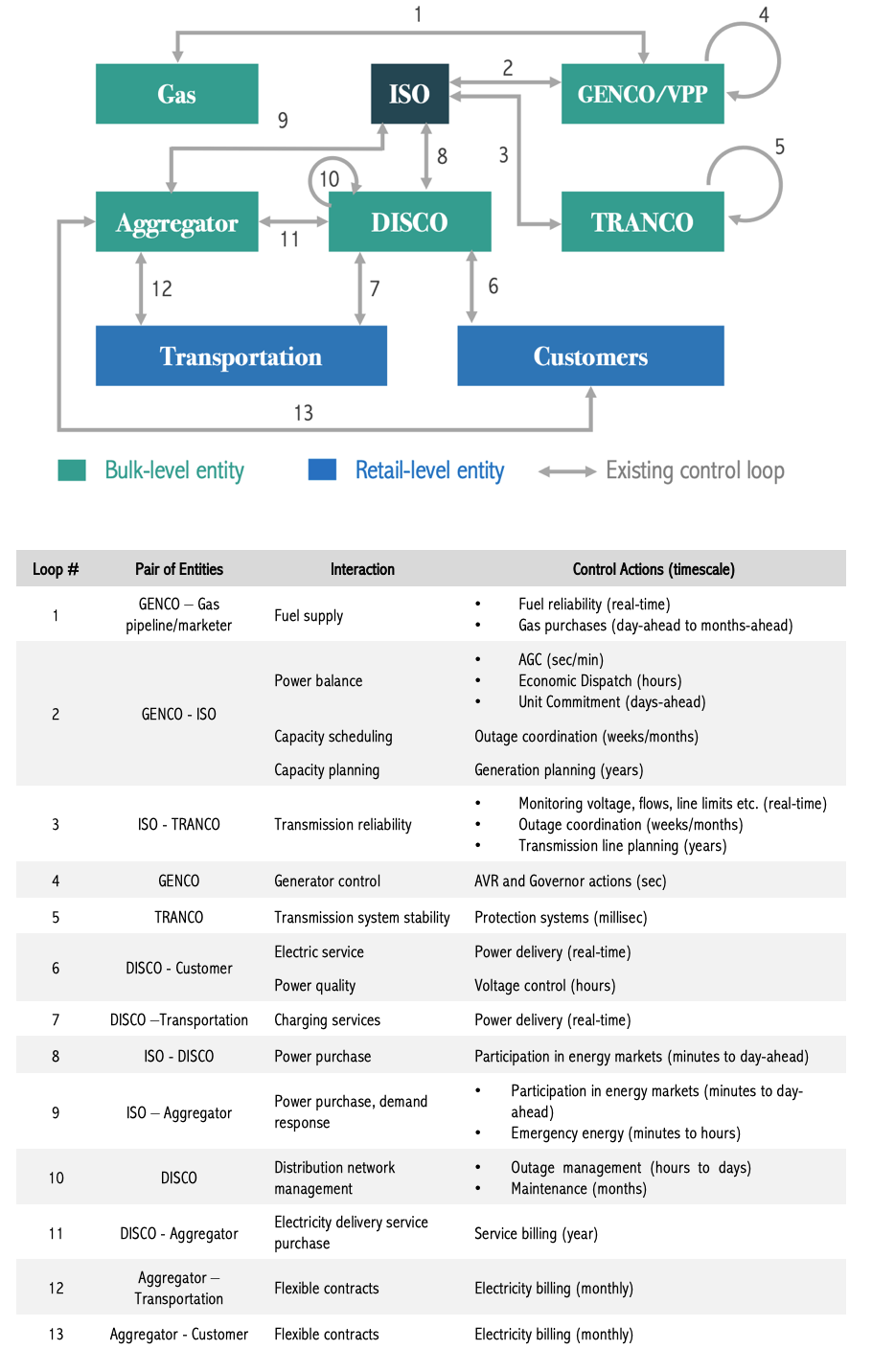}
    \caption{The existing control loops in \textcolor{black}{contemporary} power systems}
    \label{fig:old_loops}
\end{figure*}

The next subsection {\color{black}{describes}} the operational and planning aspects in transmission and distribution systems.
\subsection{Transmission/Distribution Grids: Planning and Operation} \label{sec:TandD}

\subsubsection*{Transmission System Planning} \label{sec:trans_planning}

Transmission systems in the US are planned on a regional basis with the aim of meeting future demand and maintaining grid reliability as the fleet of generating resources and their locations continues to evolve. In some areas Regional Transmission Organizations (RTOs) conduct the transmission planning process.\footnote{The terms RTO and ISO are often used interchangeably.} Transmission owners are predominantly compensated for their investments in the transmission network via cost-based rates. Building transmission infrastructure is a complicated and slow process, with concerns being raised that the process may not be able to accommodate the needs of large-scale renewable integration. The best renewable resource areas are often in more remote locations where the electric transmission infrastructure is weak \cite{ESIG2021}.

\subsubsection*{Transmission System Operations} \label{sec:trans_ops}
The ISO is responsible for running the wholesale market into which GENCOs (Loop 2 in Figure \ref{fig:old_loops}) and DISCOs (Loop 8) submit their offers and bids for power, respectively. The wholesale market run by the ISO consists of two stages: the day-ahead market and the real-time market. In turn the GENCOs are responsible for procuring fuel (Loop 1) for fulfilling the schedules determined by the ISO. The ISO has to take into consideration several other issues including how to ensure reliable delivery of power in the event of disruptions aka contingencies. The ISO also has a number of other functions needed for the reliable operation of the transmission level grid, including administering products for energy, ancillary services, capacity and financial transmission rights. The ISO also coordinates with the TRANCOs on transmission system outage scheduling and planning (Loop 3). The TRANCOs are responsible for maintenance of transmission equipment (Loop 5). In non-ISO areas the vertically integrated electric companies continue to be responsible for both transmission and distribution system operations.

\subsubsection*{Distribution System Planning} \label{sec:IRP}
Today typically the DISCO owns the distribution grid and is responsible for planning and operations to maintain its reliability. It is responsible for procuring power to supply the demand of all its customers (Loops 6 and 7). Some DISCOs generate their own power (particularly in the vertically integrated paradigm), or purchase power from other entities via power purchase agreements, or from a wholesale electricity market. 

Electric distribution companies (utilities) undertake a comprehensive process to develop Integrated Resource Plans (IRPs) to ensure that there is an adequate portfolio of supply-side and demand-side resources to meet the peak demand and energy needs of their customers at least cost, typically over a 10-20 year planning horizon \cite{wilson2013best}. Resource planning requirements differ from state to state in the US. The plans account for a variety of risks including load growth, fuel prices, fuel diversity and emissions, and require the involvement of state Public Utility Commissions (PUCs) along with other stakeholders. Thereby, societal choices are introduced into the planning of the grid infrastructure, for example, through state renewable portfolio standards.

\subsubsection*{Distribution System Operations} \label{sec:dist_ops}

Many DISCOs use Supervisory Control and Data Acquisition (SCADA) systems for ensuring the correct operation of substation equipment, through data acquisition, monitoring and control of the distribution system (Loop 10). SCADA combines hardware, software and communications to monitor various equipment, send control signals and perform key operations such as voltage control, load balancing, power quality monitoring, and fault protection.
{\color{black}{DISCOs are responsible for maintaining the voltage, peak demand, power factor and phase balance in 
the distribution system. DISCOs also attempt to ensure the continuous supply of electricity to their customers and take actions such as repair or replacement of equipment in order to restore supply after outages. Distribution Automation is playing an increasing role in maintaining the reliability of the distribution grid \cite{Ipakchi2009, madani2015distribution}. These technologies can locate and isolate faulted portions of the circuit, and automatically reconfigure the circuit to ensure fewer customers suffer outages \cite{Aboelsood2017}. Voltage management is a key function which improves energy efficiency, power factor and reliability of power distribution systems. This is accomplished by coordinating voltage regulators, load tap changers, and capacitor banks. The increase in penetration of distributed energy resources presents voltage control challenges for distribution systems \cite{sun2019review}. }}

Today most intelligence at the distribution level is in substations with limited control and visibility downstream through devices such as reclosers \textcolor{black}{and} customer smart meters. \textcolor{black}{In the future, more intelligence, such as EV software, smart thermostats, and home energy storage systems, are likely to be deployed behind the customer meters.}

\subsection{The Hierarchy of Control Loops in the Present Architecture}\label{sec:controlhierarchy}


The present architecture of power systems includes a well defined hierarchy of controls, many automatic and others manual, in order to ensure the system stays in the normal operating condition or returns quickly to the normal state when disrupted.

The stable operation of alternating current based power systems requires the monitoring and control of several key variables. In addition to real and reactive power, the main variables that need to be controlled include voltage, frequency and the phase angle of voltage. Further at the distribution level power factor and harmonics are also important. \textcolor{black}{The control loops can be described in terms of their input-output relationships.}




A few of the control loops at the transmission level are as follows:

\begin{itemize}
    {\color{black}{\item Protective relaying: A key objective of any protection system is to avoid or reduce damage to the expensive power system equipment, as well as limit the outage duration and impacted area. Transmission line relays detect faults and the faulted components are isolated from the rest of the system as quickly and as accurately as possible (Loop 5) \cite{blackburn2014protective}. Generators also need to be protected against faults and variety of abnormal operating conditions (Loop 4) \cite{1339328}. Modern high-speed relays usually operate in less than 50 ms.}}
    \item Automatic Voltage Regulators (AVRs): {\color{black}{Synchronous generators are equipped with AVRs for automatically maintaining the terminal voltage of the machine. The input to the AVR is the measured voltage, which is then compared to a reference value, and this difference or \underline{error} is fed to the control block. The voltage regulator output is used to control the excitation system of the generator, so as to cancel out the voltage error}}. This is a local and automatic control action (Loop 4), {\color{black}{with response times in the milliseconds timescale}}. {\color{black}{Proportional-integral-derivative (PID) control is a common control method, and a number of PID controller tuning approaches have been studied \cite{Panda2012, George2018}.  In addition to AVRs there are a number of supplementary controls which comprise the excitation systems of synchronous machines \cite{7553421}. The dynamic behavior of technologies such as wind turbines and solar photovoltaics is quite different from synchronous generators, and ensuring the fidelity of controller models in order to analyze their impact on
    system stability is an important area for research \cite{osti_1349211, Essallah2019}}}. 
    \item Frequency Control: {\color{black}{Maintaining
    system frequency within certain pre-specified limits is essential for power system reliability. Following a grid disturbance, the energy in the rotating masses of synchronous machines is naturally extracted, i.e., through
    inertial response. As the frequency falls,}} the first frequency control loop (aka primary or droop) is the automatic action by governors of synchronous generators (Loop 4) which sense the local frequency deviations and respond immediately to try and arrest the frequency change within seconds by changing the speed of the synchronous generator. {\color{black}{With the increasing penetration of asynchronous generation, the inertia in some systems has begun to degrade, resulting in frequency stability challenges. The solutions proposed consider emulated inertia from wind and storage systems \cite{Rezkalla2018}}}.

    Next comes the secondary frequency control (Loop 2), also called Automatic Generation Control (AGC), which ensures that frequency is brought back to the nominal value within minutes. This is done in a shared fashion by all participating generators via changing the set-points of the governors, and is coordinated centrally by the System Operator.
    {\color{black}{Imbalance between generation and demand in an area results in a deviation of the grid frequency from the nominal value (either 60 Hz or 50 Hz depending on the country). The frequency deviations of several cooperating adjacent areas can be measured and can be used as multivariable feedback to increase/decrease the set-points for generators in the areas. This control loop acts at the 30 s - 20 min time scale.}} 
    {\color{black}{Various control aspects of AGC as well as investigations into systems that include wind turbines, solar photovoltaic systems and battery storage are considered in \cite{1388528}. While historically only conventional generators have provided AGC, a number of researchers and some pilot projects have investigated the use of distributed resources such as electric vehicles and water heaters \cite{6204212, Hui2012}.}}
    \item Economic Dispatch (ED): While governor and AGC actions may be adequate to return the system to normal frequency, the result may not be economically optimal. The System Operator solves an optimization problem, leading to ``economic dispatch'' which minimizes the cost of balancing generation and load, subject to various constraints such as unit and transmission line limits. 
    ED is part of Loop 2, {\color{black}{and in US ISOs it is typically carried out every 5 minutes. The inputs to ED include the generator energy offers, measured generation outputs, unit limits, load forecast, transmission line flows and reserve requirements. The ISO clears the market resulting in prices and dispatch targets for generators. The ISO communicates these dispatch targets to the generators which then adjust their power outputs to follow the dispatch instruction. While centralized ED continues to be the mainstream paradigm, the issue of climate change and emergence of distributed energy resources have led, respectively, to research interest in environmental ED and distributed ED methods \cite{Tsegaye2019}.}}
    \item Unit Commitment (UC): {\color{black}{Unit commitment (UC) is an optimization problem used to determine the schedules of generators to balance the varying system demand at least cost, under various constraints.}} UC also forms a part of Loop 2. The inputs to the UC problem include the generator costs, demand forecast, inter-temporal constraints and various system security constraints. The {\color{blue}{outputs are}} the on/off decisions for each unit for all hours in the commitment horizon. In some cases system operators may also make \underline{out-of-market} decisions to turn on units in order to maintain system reliability. The UC process in US ISOs is implemented in multiple stages from days-ahead to hours-ahead, and uses deterministic optimization, while researchers have proposed {\color{blue}{using}} stochastic and robust approaches to deal with the increasing uncertainty due to renewables \cite{Chen2014}.  
    \item Outage Coordination (OC): System Operators coordinate maintenance schedules for both generators and transmission lines (Loops 2 and 3 respectively), weeks or months in advance in order to ensure adequate capacity will be available to meet anticipated demand, and that healthy levels of resource margins are maintained for ensuring reliability despite adverse events. {\color{black}{The inputs to the outage scheduling process are the generation and transmission outage/maintenance submissions. The output is the ISO's decision to either approve, suggest alternate dates, or deny the outage based on system reliability needs.}}
    {\color{black}{\item Long term fuel and hydro planning (FHP):  In the US most natural gas-fired generators rely on long-term \underline{firm} contracts while the rest are split between \underline{interruptible} spot market purchases only and a mix of the two types. Since firm contracts are usually more expensive while spot contracts don't guarantee fuel delivery, the generator's strategy is based on a trade-off of cost vs. risk. 
    For hydro generation, in countries like Brazil, a longer scheduling horizon may be needed due to the seasonal inflows of water into the reservoirs \cite{Helseth2020}. For nuclear power plants their refueling related outages are scheduled during seasons with low electricity demand. In the US the nuclear power plants typically refuel every 18 to 24 months.}} {\color{black}{The initial preparation for a refueling outage may begin as much as 3-5 years in advance, and on average the outage duration is about a month.}}
    \item Infrastructure Planning and Construction (IPC): The physical grid infrastructure is planned based on a time horizon of years or decades. This is where decisions on building new plants (Loops 2 and 4) or transmission lines (Loops 3 and 5) are made. {\color{black}{In the US many large utilities use integrated resource plans to determine their generation expansion over a multi-year horizon. The inputs include expected demand growth, policy goals, constraints, and resource options. The output is the most cost effective mix of both supply and demand side resources to ensure reliable service to customers.}}
\end{itemize}

A few of the control loops at the distribution level are as follows:

\begin{itemize}
\item Power Quality/Reliability: DISCOs perform a number of control actions on the distribution grid components in order to provide reliable power to Customers (Loop 6). These include {\color{black}{distribution/feeder}} reconfiguration, power factor correction, {\color{black}{and voltage regulation}}. {\color{black}{The distribution reconfiguration aims to reduce distribution line loss by changing the topology of the distribution system. The input to the distribution reconfiguration process is the load profile, and its output is the list of on/off statuses of switches \cite{6867399}. The power factor correction can be achieved by single/three phase voltage source inverters (VSIs) \cite{847283}. The control loop of the VSI takes measurements of the terminal voltage and load current as inputs, and generates control commands so as to inject the specified current that can improve the power factor \cite{847283}. The voltage regulation in conventional distribution systems is accomplished by shunt capacitor banks and on-load tap changers (OLTCs) \cite{6942276}. The voltage regulation process determines the on/off statuses of the capacitor banks and the tap settings of the OLTCs by observing voltage measurements.

Additionally, there are a number of other control or switching actions that may be undertaken to protect key equipment from abnormal conditions such as faults, over-voltage, over-heating, and abnormal frequencies.}}
\item Power Purchases: In order to provide adequate power to their Customers, DISCOs may participate in wholesale electricity markets operated by the ISOs (Loop 8). \textcolor{black}{Based on the load forecast and other market information, DISCOs determine the amount and price of electricity that they need to trade.}
\item Demand Response: Aggregators may offer demand response services, i.e., offering to reduce demand in ISO markets on behalf of multiple customers (Loop 9). \textcolor{black}{Depending on various incentives provided by the aggregators (e.g., electricity price), \textcolor{black}{either} the electricity end users or the load management software determine\textcolor{black}{s} the amount of electricity usage during different periods of time, manually or automatically, \textcolor{black}{respectively}.}
\item Electricity Billing: If a region has no electricity retail market (i.e., retail choice), the DISCO \textcolor{black}{will be the sole provider and} charge customers for their electricity usage (Loops 6 and 7). \textcolor{black}{In such a scenario usually the electricity tariff is the result of a ``rate case'' process, in which the inputs are the required revenue, operating costs and allowed rate of return. The rate case is approved by the appropriate regulatory body.} In a region with a retail market, aggregators charge customers based on the rate in the contracts that their customers choose (Loops 12 and 13). \textcolor{black}{The contracts provided by the aggregators specify different electricity usage patterns.} The aggregators also need to pay their local DISCO for the power delivery services provided by the DISCO (Loop 11).

\end{itemize}



The development of the control laws and their implementation in software has allowed for the building of the largest human-made engineering system of the 20th century, namely the electrical grid.

\section{The Drivers of Change}\label{Sec:change}

\begin{figure}[tbh]
    \centering
    \includegraphics[width = \linewidth]{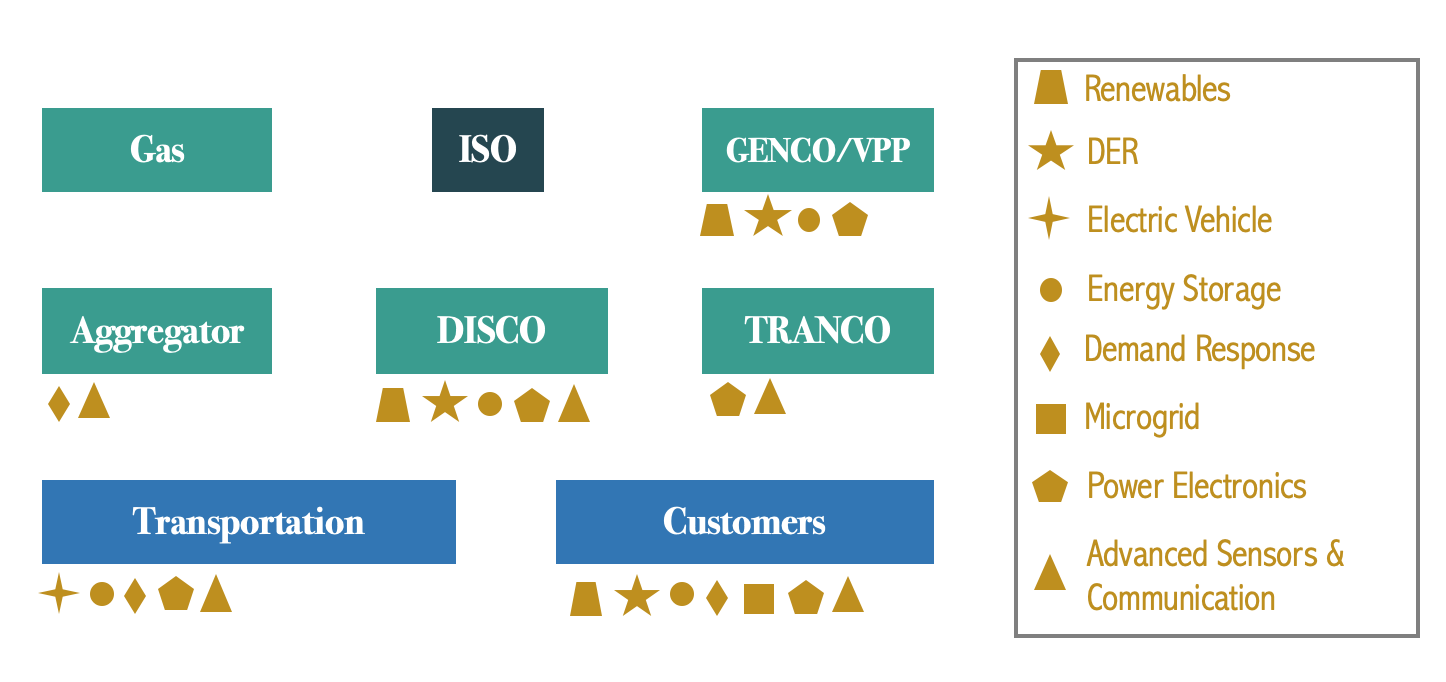}
    \caption{The key drivers of change of future electric energy systems }
    \label{fig:enablers}
\end{figure}

While the present architecture of the electric energy system has served the electricity industry well over the decades, \textcolor{black}{it can be anticipated that} there will be a major paradigm shift in the way electricity gets generated, delivered, and consumed. This is driven by the societal choice of decarbonization which will have two major ramifications on the electricity sector, first, a substantial number of \textcolor{black}{conventional} generating units will be replaced by renewable energy resources. Second, there will be a substantial amount of new load,
a prime example being the electrification of transportation \cite{jenkins2021mission,stoddard2021three}.
When generation becomes uncontrollable, as in the case of renewables, the balance between generation and demand
will need to be maintained by incorporating storage to buffer imbalances between the two, and by changes in how demand is managed or controlled. The key drivers of changes among different entities are shown in Figure \ref{fig:enablers}. Next, these key drivers \textcolor{black}{are elaborated upon}.

\subsection{Integration of Variable Renewable Energy} \label{sec:VRE}
Over the past two decades wind and solar resources have been increasingly integrated at the transmission level. The  inherent variability of these resources presents new challenges to the operation of the grid. Reliable grid operation will need more operational flexibility to balance the intermittent renewable generation.
One example of the challenges posed by renewables is illustrated by the now infamous ``Duck Curve" in California \cite{CAISORep,9585313}, where the net load is low in the middle of the day since solar production is at its highest and then rises rapidly towards the evening peak as demand increases and sunlight fades. This results in the need to ramp the generation fleet by a significant amount over a span of just a few hours, posing a major challenge.

\subsection{Aggregation of Distributed Energy Resources (DERs) in Wholesale Markets} \label{sec:ferc2222}

While historically the spatial separation has resulted in the distribution system being characterized as purely load from the transmission level perspective, the paradigm has begun to change to one where transmission and distribution system operations will become more integrated. A key reason behind this is the increasing deployment of Distributed Energy Resources (DERs).

In 2020 \textcolor{black}{the} Federal Energy Regulatory Commission (FERC) issued Order 2222 to allow aggregations of DERs to participate in wholesale electricity markets \cite{ferc2222}. This will allow DERs to provide energy, ancillary services (such as frequency regulation and contingency reserves)  and capacity, and compete in wholesale markets with large generators. This order permits aggregators to pool multiple small DERs so as to meet minimum size thresholds and other requirements of the ISOs. The current participation models, market clearing algorithms  and other mechanisms  will need to be modified in order to account for the suite of technologies that might be included in DERs, including electric storage, distributed generation (DG), demand response, energy efficiency, thermal storage and electric vehicles. Currently only two ISOs in the US, namely New York (NYISO) and California (CAISO) have DER aggregation programs.

At present, ISOs do not have visibility into the distribution grid, but this could be a concern when large numbers of DERs start participating in the wholesale market via aggregators. When the DER aggregator tries to maximize its profit in the wholesale markets it will have an impact on the operation of the distribution grid. It is important to ensure that the participation of DERs in the wholesale market does not violate any of the distribution system constraints. Thus, there is a need for a coordination framework to exchange information and control signals between the aggregator, the DISCO and the ISO. 

Researchers are looking into the question of whether there is a need for a new entity called ``Distribution System Operator" (DSO) to coordinate retail transactions and operations, analogous to the way the ISO coordinates these at the wholesale level. A distribution level market may be able to account for distribution constraints (such as voltage and line flow limits) and provide appropriate price signals through \textcolor{black}{distribution} Locational Marginal Prices (dLMPs) \cite{haider2020toward}. For instance, Andrianesis and Caramanis \cite{andrianesis2020distribution} presents an approach to consider the impact of DER scheduling on distribution transformer degradation.




\subsection{Electric Vehicles} \label{sec:EV}
One of the largest sources of emissions is the transportation sector. Hence a transition from the internal combustion engine to electric vehicles is necessary (but not sufficient) for achieving decarbonization goals. To accomplish these there is a need for significant investments in EV charging infrastructure. Further, uncontrolled charging of EVs could result in serious challenges as the peak demand could far outstrip what the distribution equipment is rated for \cite{mckinsey2018}.

\subsection{Energy Storage} \label{sec:storage}
Historically the grid has operated largely on a just-in-time principle where electricity generation attempts to balance load on a second-by-second basis. Hence energy storage technologies could be important to effectively integrating intermittent renewables into the grid. Pumped Hydroelectric Storage (PHS) currently makes up 96\% of worldwide energy storage capacity \cite{DOE2020storagereview}. Other technologies include batteries, molten salt (thermal storage), compressed air and flywheels. Energy storage can provide a number of benefits to the grid including capacity firming, peak demand management, ancillary services and transmission and distribution upgrade deferral. However, there are a number of challenges with regards to scaling these solutions. For instance, with PHS there are only a few locations where the reservoirs can be located. For other technologies, such as batteries, the capital costs are still quite high \cite{Shaqsi2020}. Further, there are sustainability issues as the mining of lithium for lithium-ion (Li-ion) batteries has serious environmental impact.

\subsection{Demand Response} \label{sec:DR}
Demand response involves the modification of energy usage by reducing or shifting it in time in order to better match the supply.
This could be in the form of direct load control where utilities can cycle appliances such as air-conditioners and water heaters in exchange for a financial incentive in order to manage peak demand. Another approach to elicit customer participation via financial incentives is to provide them with time-based rates. These include time-of-use pricing, critical peak pricing, variable peak pricing, real time pricing, and critical peak rebates. Smart meters are a key enabling technology for such programs. Time-of-use rates could be the key to managing the increased peak demand arising due to the charging needs of EVs \cite{mckinsey2018}. \textcolor{black}{Data-driven approaches can be used to support key decision making processes in demand response programs, e.g., customer behavior modeling and control \cite{9304500}, and electricity baseline prediction \cite{ming2020prediction}, due to a lack of physical-based models.}

\subsection{Microgrids} \label{sec:microgrids}
Microgrids are local grids which can be formed by connecting a group of loads and distributed energy resources, and which can disconnect from the traditional grid to operate autonomously.
They are useful for serving remote areas, or  when the main grid suffers disruptions such as due to storms. Hence microgrids may help to improve grid resilience. In the future electric energy systems, microgrids may be used more widely, particularly as the deployment of DERs and grid-edge intelligence increases. 


\subsection{Power Electronics} \label{sec:electronics}
A number of the newer energy technologies such as solar panels, wind generators and batteries interface with the grid via power electronics inverters. As \textcolor{black}{society} transitions away from a grid with large synchronous generators towards one with such Inverter-based Resources (IBRs) there will be a number of challenges, particularly since those synchronous machines have provided much of the controls (both frequency and voltage) needed to maintain grid stability. One key challenge will be managing the grid with lower inertia, since the lack of inertia will impact grid stability. There is a need to research alternative methods known as grid-forming controls which can exploit the capabilities of the IBRs \cite{lin2020research}.

\subsection{Advanced Sensors and Communication} \label{sec:sensors}
Over the past few decades a number of smart grid technologies have begun to be deployed that provide improved capabilities to manage the grid through more granular measurements (both spatial and temporal), communications and intelligent controls. A few examples of these include:
\begin{enumerate}
\item Advanced sensors known as Phasor Measurement Units (PMUs) that allow operators to assess transmission grid stability.
\item Sensors that measure transmission line capacity in real time thereby enabling dynamic line rating changes.
\item Smart sensor technologies at distribution level, such as Fault Circuit Indicators (FCIs), that give greater visibility and control below the substation.
\end{enumerate}

These new sensors, together with new communication capabilities such as 5G, offer opportunities for designing new control loops around the future electric energy systems to achieve higher reliability and efficiency. \textcolor{black}{The data collected by the advanced sensor infrastructure will allow for the employment of data-driven approaches to address grid operational tasks, such as grid monitoring \cite{9043670,9253523,8362302}, control \cite{8873679} and protection \cite{9029268}. The domain-tailored artificial intelligence leveraging the grid operational data can greatly complement the conventional grid operation schemes that heavily depend on physics-based models \cite{9091534}.}

\section{Research Opportunities and Challenges of \textcolor{black}{Information and} Control Architecture in Future Power Grids}\label{Sec:Challenges}


Addressing all the issues raised by incorporating the new entities as well as the new capabilities described above poses several fundamental challenges for the electric energy sector,
from operations to planning, and from transmission to distribution systems.
Any architectural changes will need to satisfy several objectives:
1) They continue to satisfy the primary objective of providing reliable, efficient, and low-carbon electricity; 2) They are well aligned with the responsibilities and interests of all the new and old entities; 3) The new mechanisms are backward compatible in view of the massive \textcolor{black}{existing} investment in infrastructure; and 4) They do not constrain future innovation, but encourage it.
\textcolor{black}{Specifically, the architecture design for the power grids should also address the following key research questions:}


\begin{itemize}
    \item R1: What is an architectural design for operating networked microgrids, and for designing protocols that allow for plug-and-play with security certificates? 
    \item R2: How to control and operate energy storage devices across transmission and distribution systems with deep penetration of variable resources?
    \item R3: What are the control protocols that could incorporate massive amounts of inverter-based resources (IBRs) (e.g. wind, solar, and EV charging) into the existing electric grid at scale? 
    \item R4: What kinds of security guarantee\textcolor{black}{s} can be given in the presence of many smart controllers and sensors across the electric grid? 
    \item R5: What is a 
    coordinated scheduling and planning 
    framework that would allow for deep penetration of vehicle electrification?
    \item R6: What are new utility business models and rate structures to incentivize innovation and investments in grid-edge technologies? 
    \item R7: What are appropriate market mechanisms to dispatch massive amounts of uncertain resources with a priori risk awareness? 
    
    \item R8:  How to manage a multi-energy system with stronger coupling between gas and electricity?


\end{itemize}

{\color{black}{

Traditionally power systems have relied on a mix of centralized controls (e.g., scheduling, dispatch) and decentralized controls (e.g., protection relays, AVRs, Governors).
As the industry is in transition with the increase in penetration of DERs, it is likely to see some grid support services being provided by controllable resources, usually with very similar \textcolor{black}{information and} control architectures, as initially many end users will continue to depend on the bulk power system for at least part of their power needs. Further, there continue to remain economics of scale e.g. large on-shore and off-shore wind farms, which could justify the need for continuing some centralization of control/coordination. There may be fully or mostly decentralized schemes in niche cases, such as in islands and remote micro-grids. While decentralized control is generally considered faster than centralized, there may still be a need for coordination/communication. With independent micro-grids we may see peer-to-peer transactions with distributed intelligence to maintain grid reliability.}}

In outlining an architecture for the future electric energy system, \textcolor{black}{this paper} has adopted a design philosophy that
 decisions should be delegated to the lowest, i.e., most local, level that is appropriate.
 This is motivated by the consideration of spurring  innovation at the edge. However, where necessary, higher level coordination \textcolor{black}{is advocated}.
As an example, \textcolor{black}{one lesson learned from the Texas 2021 Power Crisis is that} a tighter and high level integrated joint design of the gas network and the electric grid \textcolor{black}{is desirable}, as opposed to \textcolor{black}{the} current distributed decision making \cite{Texas2021}.


Building upon these new technologies as well as the design philosophy, \textcolor{black}{this paper} outlines additions to the \textcolor{black}{information and} control architecture
of the electric grid that would allow for large scale integration of renewables, but would still be backward compatible. 
Several new control loops at the transmission and distribution levels \textcolor{black}{are summarized in Figure \ref{fig:new_loops} and presented} in the next two sections, respectively. 

\section{New Loops in Future Transmission Systems}\label{Sec:NewLoopsTran}

\begin{figure*}
    \centering
    \includegraphics[width = 5in]{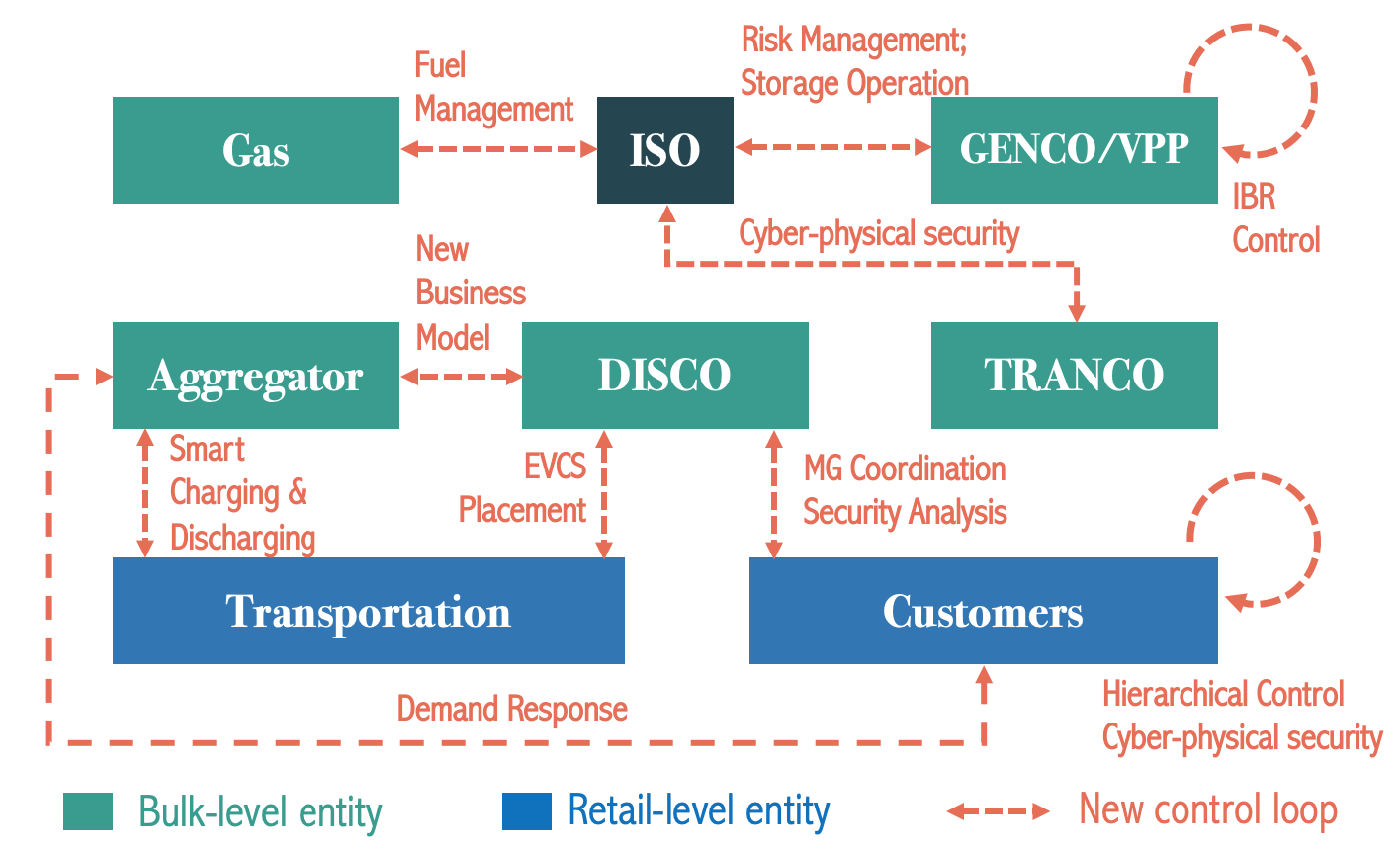}
    \caption{New control loops for future electric energy systems with deepened renewable energy penetration}
    \label{fig:new_loops}
\end{figure*}

Given the technological innovations and societal objectives of decarbonization and electrification, the \textcolor{black}{information and} control architecture of the future electric energy systems will need to be adapted to best accommodate the needs. These new control loops range from integrating individual resources to resource-grid coordination and to cross-infrastructure coordination. Some of the new control loops \textcolor{black}{are summarized in Figure \ref{fig:new_loops} and they are presented} in the following subsections.


\subsection{Inverter-based Resources Control} \label{sec:IBRctrl}

\begin{figure}[htb]
    \centering
    \includegraphics[width = \linewidth]{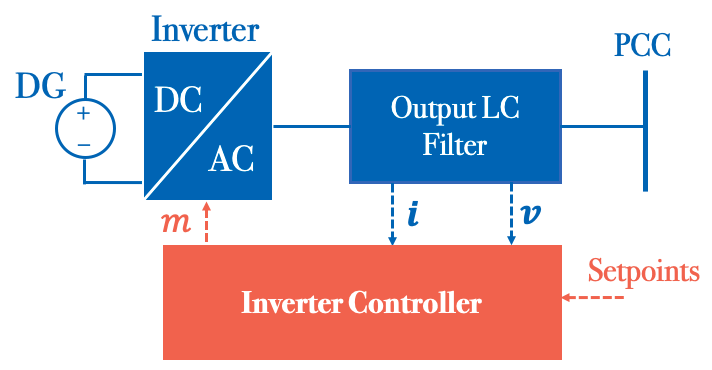}
    \caption{A grid-connected inverter with a controller at the grid interface}
    \label{fig:grid_connected_inverter}
\end{figure}

Inverters are the power electronics devices that can convert Direct Current (DC) to Alternating Current (AC) electricity.
They are becoming \textcolor{black}{part of the} \textit{key \textcolor{black}{grid} equipment} since a number of DG technologies such as solar, wind power and batteries need to convert DC electricity to AC in order to connect to the bulk power system.
Figure \ref{fig:grid_connected_inverter} shows a DC energy resource connected to the grid at a \textcolor{black}{Point of Common Coupling} (PCC) via an inverter and a low-pass filter. The inverter is regulated by a feedback controller that observes the inverter terminal voltage and current, and tunes the inverter modulation index, in order to track a given setpoint. Almost all inverters currently connected to the grid are grid-following inverters. The setpoints for the grid-following inverter are real and reactive power which can be determined by solving the Optimal Power Flow (OPF) problem \cite{6805674,6417004,SU201245}. A Phase Lock Loop (PLL), a current controller, and a Pulse Width Modulation (PWM) module are three typical functional blocks in a grid-following inverter \cite{9528341}. 
The grid-following inverter is designed for the situation where the frequency and voltage magnitude at the PCC are in a nominal range. This type of inverter cannot provide a nominal frequency and a voltage magnitude at the PCC autonomously without a well-established grid. As a result, grid-following inverter\textcolor{black}{s} \textcolor{black}{by themselves} cannot form an autonomous power grid in the restoration stages. This motivates the development of a new type of inverter called {\textit{grid-forming inverter}} which is described next.

Grid-forming inverters can regulate terminal voltage magnitude and frequency at the PCC by tracking the frequency and voltage setpoints, and can provide restoration support \cite{9528341,lin2020research}. Grid-forming inverter controllers can be broken down into three categories \cite{lin2020research}: 1) droop controllers \cite{4118327}, 2) synchronverter controllers \cite{beck2007virtual,5456209}, and 3) virtual oscillator controllers \cite{7317584}. Since there is a tight coupling between grid frequencies (voltage magnitudes) and real (reactive) power in a power grid with small line losses, the {\textit{droop controllers}} stabilize grid frequencies (voltage magnitudes) by changing \textcolor{black}{the} real (reactive) power that DGs inject into the grid \cite{195899,4118327}.
They can also determine how much real and reactive power each DG contributes in order to eliminate real and reactive power imbalance. A droop controller typically consists of a power controller, a voltage controller, \textcolor{black}{a} current controller, and a PWM module \cite{9528341,4118327}. A {\textit{synchronverter controller}} enables an inverter to mimic the behavior of a synchronous machine \cite{beck2007virtual,5456209}. Such a control strategy is motivated by the fact that the contemporary grid apparatus, e.g., protection and control, is designed for a grid where synchronous generators are dominant. Engineering an inverter-based resource that behaves like a synchronous generator makes it easier for it to be integrated into an existing power grid, since its host grid does not require many changes. Another attractive feature of the synchronverters is that they can provide their host grid with inertia, which makes the grid less sensitive to disturbances. \textcolor{black}{Techniques for grid inertia estimation and improvement are reported in \cite{8708244,9699083} and the references cited therein.} \textcolor{black}{A {\textit{virtual oscillator controller}} enables a converter to mimic a limit-cycle oscillator such that the converter can be synchronized with other converters.}
The parameter tuning procedure of a virtual oscillator controller can be achieved in a systematic way \cite{7317584}.

There are decades of experience in operating the grid with large synchronous generators, with well-developed control loops (discussed in Section \ref{sec:controlhierarchy}) based on well understood models and interactions. However, as \textcolor{black}{the electricity supply systems} transition to a grid dominated by IBRs, \textcolor{black}{there is a} lack of experience and body of research to operate the grid reliably. The key control challenges include inertia, frequency regulation, voltage control, fault protection schemes, black start and wide-area coordination. It is also important to recognize that in many systems there will be a period of transition where both the conventional and new resources will have to operate in coordination.

\subsection{Cyber-Physical Security of Control Loops} \label{sec:cps}
Decision making in \textcolor{black}{the} operation of bulk power systems heavily relies on massive distributed measurements. For example, the electricity market operation is built upon a wide-area state estimator that takes SCADA data as inputs.
The AGC changes the setpoints of generators based on the SCADA measurements that may be hundreds of miles away from the control center where the AGC is located. Due to the tight dependence of operational decisions on measurements, there arises a concern that a malicious adversary may compromise the grid operation by manipulating the physically distributed, poorly-secured sensors. Reference \cite{xie2011integrity} envisions a scenario where an attacker can be benefited financially by deliberately tampering with the SCADA measurements. Reference \cite{8345676} suggests that the grid can be destabilized by an attacker who intrudes into the frequency/tie-line flow measurements of the AGC system. Moreover, as the scale of IBRs grows, so does the need for communication in order to monitor and control them. Thus increasingly the inverters might become targets for cyber attacks. An attacker might use an inverter's software as an entry point for a malware attack that could spread to the rest of the grid and cause cascading failures. 
Thus there is a need for certification standards and programs that can help the industry to manage the cybersecurity risks \cite{hupp2021cybersecurity}.

\begin{figure}
    \centering
    \includegraphics[width = 3.5in]{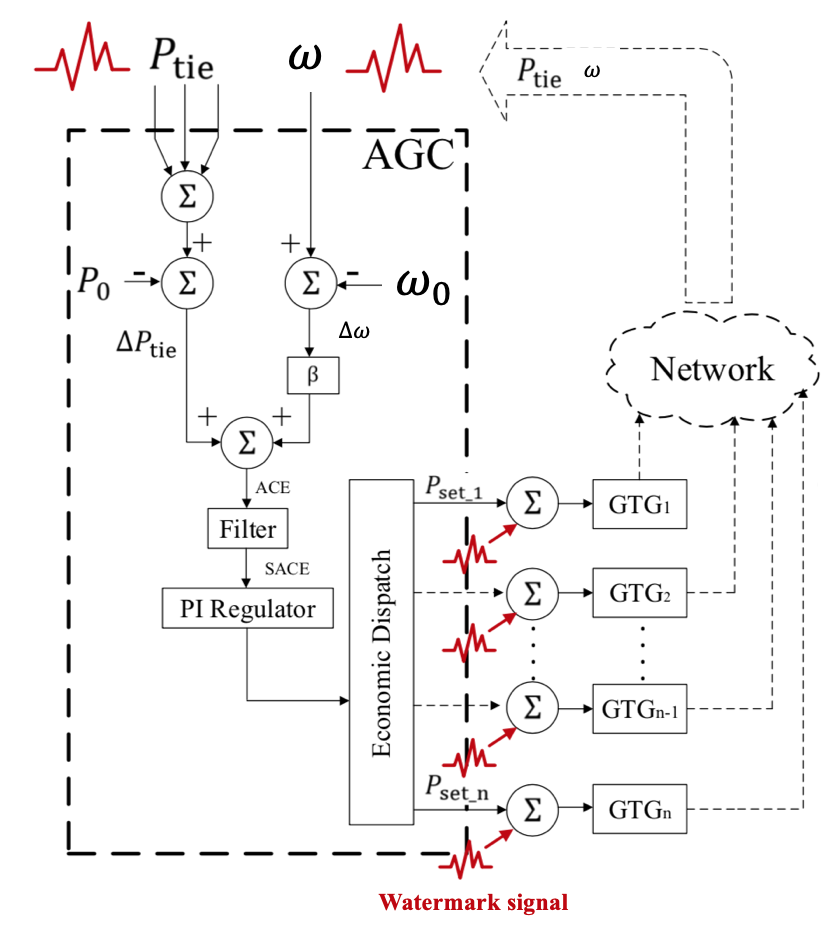}
    \caption{The dynamic-watermarking approach of \cite{satchidanandan2016dynamic} applied for securing the AGC of a bulk transmission system \cite{8345676}. A small secret noise, called ``watermark", is superimposed on the control command issued by the AGC, i.e., the power setpoint of a generator under the AGC control. The secret signal, suitably transformed is expected to be reflected in the distributed measurements feeding the AGC, i.e., measurements of frequency $\omega$ and tie-line flows $P_{\text{tie}}$. If it is not, then that indicates an attack (or other malfunction). A wide range of cyber attacks can be so detected with rigorous theoretical guarantees. ACE: area control error. SACE: smoothed ACE. PI: proportional-integral. GTG: governor turbine generator. (Source: Fig. 2 of reference \cite{8345676} \copyright IEEE 2018)}
    \label{fig:AGC}
\end{figure}

One challenge of designing a cybersecurity solution is how to distinguish cyber attacks from physical disturbances. For example, some cybersecurity solutions (e.g., \cite{8345676}) are based on power grid models that are only valid around certain operating conditions. However, large disturbances, e.g., line tripping, might drive the grid to an operating condition where the model used for the cyber solutions is not valid anymore, thereby triggering false alarms. Besides, distinguishing cyber and physical anomalies is important because different types of anomalies require different mitigation actions. Therefore, to avoid false alarms due to physical anomalies, an anomaly classification procedure is necessary for developing cybersecurity solutions. While a substantial body of literature is devoted to address the cyber-physical security of power grids \cite{song2018intelligent,9167203, mohan2020comprehensive, dibaji2019systems}, the deployment of cyber-physical security solutions in real-world power systems is still in a nascent stage. Recently, the dynamic watermarking method \cite{satchidanandan2016dynamic} has been used to detect malicious attacks on the sensors in prototypical microgrids \cite{9528341}, and tested via simulation on attacks on the AGC loop \cite{8345676} (See Figure \ref{fig:AGC}). Without guarantees on cyber-physical security, new technologies that leverage wide-area measurements will not find acceptance for real-time grid operations. 


\subsection{Control of Energy Storage}

Energy storage techniques used in power grids typically include Battery Energy Storage (BES), supercapacitor storage, flywheel storage, superconducting magnetic energy storage, and pumped hydro-storage \cite{6922604}. Reference \cite{6922604} introduces the past and present of these storage techniques.
Energy storage can play a part in providing grid operation services, such as frequency regulation, and peak shaving, which aims to eliminate short-term demand spikes that increase grid operation costs and jeopardize grid security. 
Taking BES as an example, 
some ISOs and RTOs have established a pay-for-performance regulation market where BES participants can obtain financial rewards by providing charging and discharging services in a timely manner \cite{8383984}. 
Strategies of using BES for grid frequency regulation are proposed in \cite{8383984,5109519,8935195,zecchino2021optimal}, while
\cite{8027056} proposes a scheme that leverages BES for both peak shaving and frequency regulation.

As the costs of storage technologies continue to decrease and \textcolor{black}{the} penetration of variable renewables grows, there is a need to consider the potential of energy storage in all aspects of the grid from planning to operations. Already \textcolor{black}{there has been} tremendous interest from developers in terms of the generation queues of ISOs where battery storage is combined with other technologies (usually solar) to form Hybrid Power Plants \cite{eia2021battery}.




\subsection{Gas and Electric Coordination}
\label{sec:gas_elec}

With the retirement of a number of coal generators and the overall penetration of renewables still being low, the electrical grid in the US has begun to rely more on gas fired generation. Hence the coordination of gas and electricity resources is taking on a greater importance in the reliable operation of the electricity system.
The pipeline operators and electrical generator operators need to coordinate to minimize natural gas pipeline outages during peak electric demand periods and prioritize supply to generators during cold weather when the load tends to be high due to heating needs. Further, the increase in \textcolor{black}{the installed capacity of} renewables may lead to additional strain on natural gas infrastructure to \textcolor{black}{provide adequate gas to support system ramping needs,} due to the variability \textcolor{black}{of renewable resources}.

Currently the gas and electric systems are modeled and optimized separately. Co-optimization of the electricity and natural gas operations could be more efficient than separate optimizations \cite{pambour2018value}. To improve reliability more information sharing and improved modeling capabilities are needed. In particular, as system dynamics become more important, steady-state models may not be sufficient to address the challenges. 


\subsection{Risk-aware Dispatch and Scheduling}

UC and ED are two key problems that ISOs must solve in order to maintain the steady-state reliable operation of the electric grid while ensuring \textcolor{black}{the} lowest cost for customers. Some generators require on the order of several hours to start-up or shut-down, hence the ``unit commitment problem'' of determining the schedule for the on/off decisions of different generators, is typically done one day in advance. Also, generators submit their offers on an hour-by-hour basis for the next day, and the System Operator decides on which bids to accept to meet aggregated demand bids, based on lowest cost, i.e., day-ahead economic dispatch. Further, on the given day, generators are economically dispatched, typically, on a 5-minute basis, so that the System Operator can balance generation and supply, i.e., real-time economic dispatch.
Currently these problems are formulated as deterministic optimization problems. To manage uncertainties ISOs use reserve requirements and procure headroom capacity across their commitment and dispatch processes \cite{Chen2014}. However, these requirements are relatively static and do not vary dynamically depending on the immediate system needs. As the uncertainties grow in both scale and complexity due to a variety of factors such as the increased penetration of variable renewables, more frequent severe weather events and the emergence of DERs, the traditional deterministic approach may result in an excessive need for such reserve capacity with associated increased inefficiency. Hence there is a need for approaches that explicitly consider risk and manage the large number of uncertainty scenarios optimally. 

A number of approaches have been studied to deal with the increasing uncertainty in power systems.
These include stochastic optimization \cite{papavasiliou2013multiarea, Tuohy2009}, robust optimization \cite{lorca2016multistage, thatte2014robust}, chance constrained optimization \cite{GENG2019341}, and the scenario approach-based optimization \cite{geng2021computing,ming2017scenario,modarresi2018scenario}. 
The software, hardware and modeling techniques used by the ISOs will need to be improved in order to handle both the massive increase in computational complexity as well as the variety of new generation technologies that will emerge.
While a variety of approaches are possible, it is clear that there needs to be better control of the increased uncertainties that will be present in the future.

\section{New Loops in Future Distribution Systems}\label{Sec:NewLoopDist}
This section proposes several new control loops in future distribution systems. The large-scale penetration of DERs in distribution systems adds unprecedented operational complexity which may require the distribution systems to shift towards a microgrid-based architecture. 
This constitutes a move towards a less centralized and more distributed design and operation of distribution systems. Transportation electrification introduces interdependence among utilities, aggregators, and EVs and their charging stations. The increasing interdependency 
would in turn imply a more coupled \textcolor{black}{information and} control architecture where the temporal and spatial scale separation will need to be evaluated more adaptively. 
Grid edge intelligence (e.g., smart meters, and IoTs) enables customers to play a significant part in enhancing grid operation efficiency. Further, the heterogeneous nature of load and the decreasing amount of energy transactions over distribution lines motivate changes to the business models of utilities. The above aspects can be addressed by introducing some new control loops \textcolor{black}{in Figure \ref{fig:new_loops}} into future distribution systems.  \textcolor{black}{As noted in Section \ref{Section1}, the inputs, outputs and purposes of the control loops are described, but the specific control methodologies to be used for their design, of which there are usually several, are left as topics for future research.}


\subsection{Microgrid-based Distribution Systems}
This subsection introduces the self loops at the Customers block and the loops between the DISCO blocks and the Customers block in Figure \ref{fig:new_loops}. These new loops are motivated by the increasing penetration of DERs in the distribution grids, and they are enabled by the emerging computation and communication capacities at the grid edges. \textcolor{black}{A physical architecture of future distribution systems that is fundamentally different from today's distribution grids is presented. \textcolor{black}{T}hen a hierarchical control scheme that manages the physical architecture is introduced. \textcolor{black}{F}inally tools that can be leveraged to analyze the security of the microgrid-based distribution system are described.}

\subsubsection{Physical Architecture of a Microgrid-based Distribution System} 
\begin{figure}[htb]
    \centering
    \includegraphics[width = \linewidth]{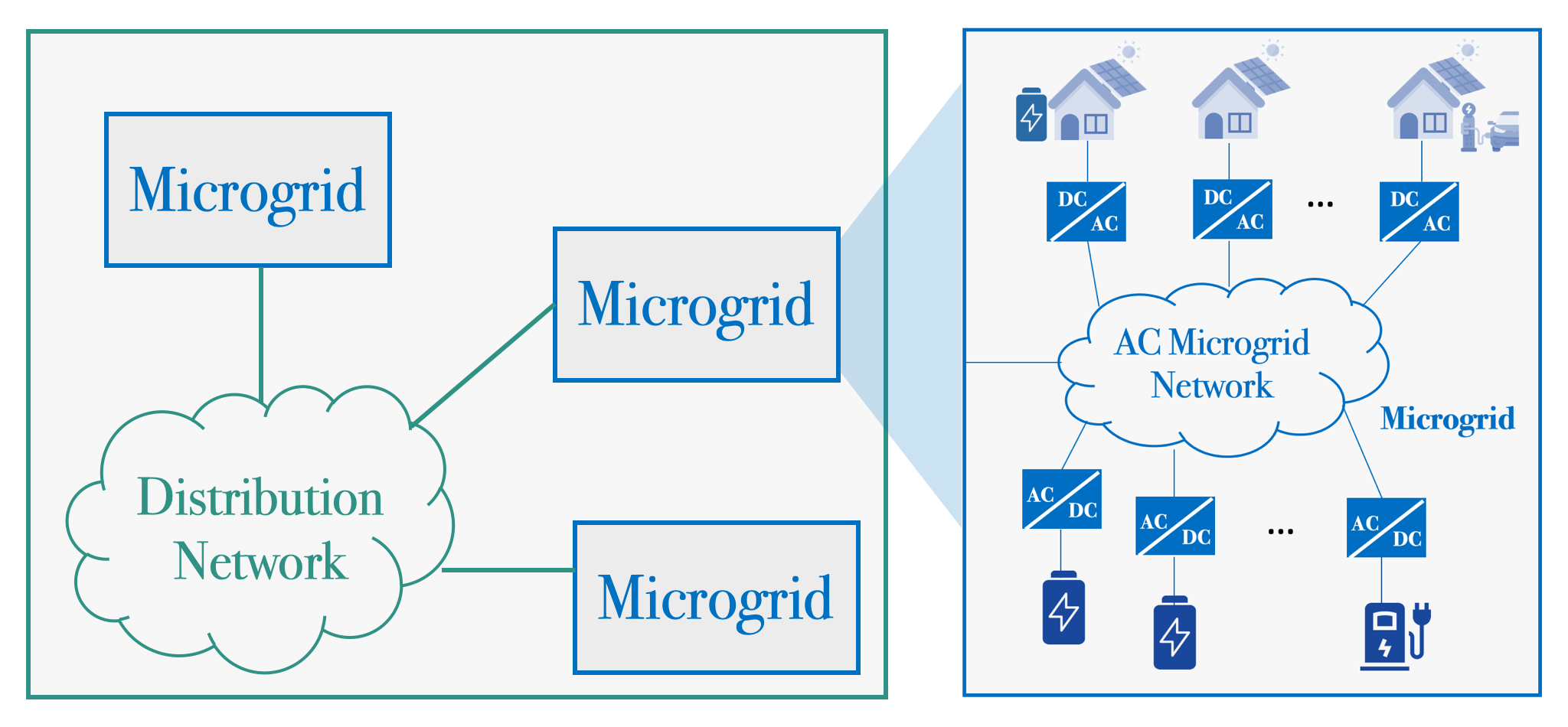}
    \caption{Physical architecture of a microgrid-based distribution system}
    \label{fig:MG}
\end{figure}

The increasing penetration of DERs poses significant challenges for distribution system operation. A future distribution system may contain hundreds to thousands of DERs (e.g., rooftop solar panels, and energy storage). The key challenge for DISCOs is how to manage such a massive deployment of DERs.
One possible physical architecture for a future distribution system able to manage massive DERs is shown in
Figure \ref{fig:MG}. The grid edge components, i.e., loads, DERs, and energy storage, are clustered into several groups, and each group of  edge components is served by a small-scale, autonomous power system, i.e., a microgrid. Each microgrid is managed by its own Microgrid Management System  ($\mu$MS). With such a configuration, the DISCO only needs to manage several $\mu$MSs, instead of controlling massive DERs. As a result, the operational complexity for the DISCO is significantly reduced. The resilience of the distribution system to natural disasters can also be greatly enhanced by a microgrid-based configuration. Each microgrid can either connect to the main grid or disconnect from the main grid and operate autonomously. Were a natural disaster to occur that caused the host distribution networks to lose their power delivery function, then each microgrid could proactively disconnect from its host grid and supply its load autonomously. Thereby, large-scale blackouts could be avoided.
\subsubsection{Microgrid Hierarchical Control}
\begin{figure}
    \centering
    \includegraphics[width = \linewidth]{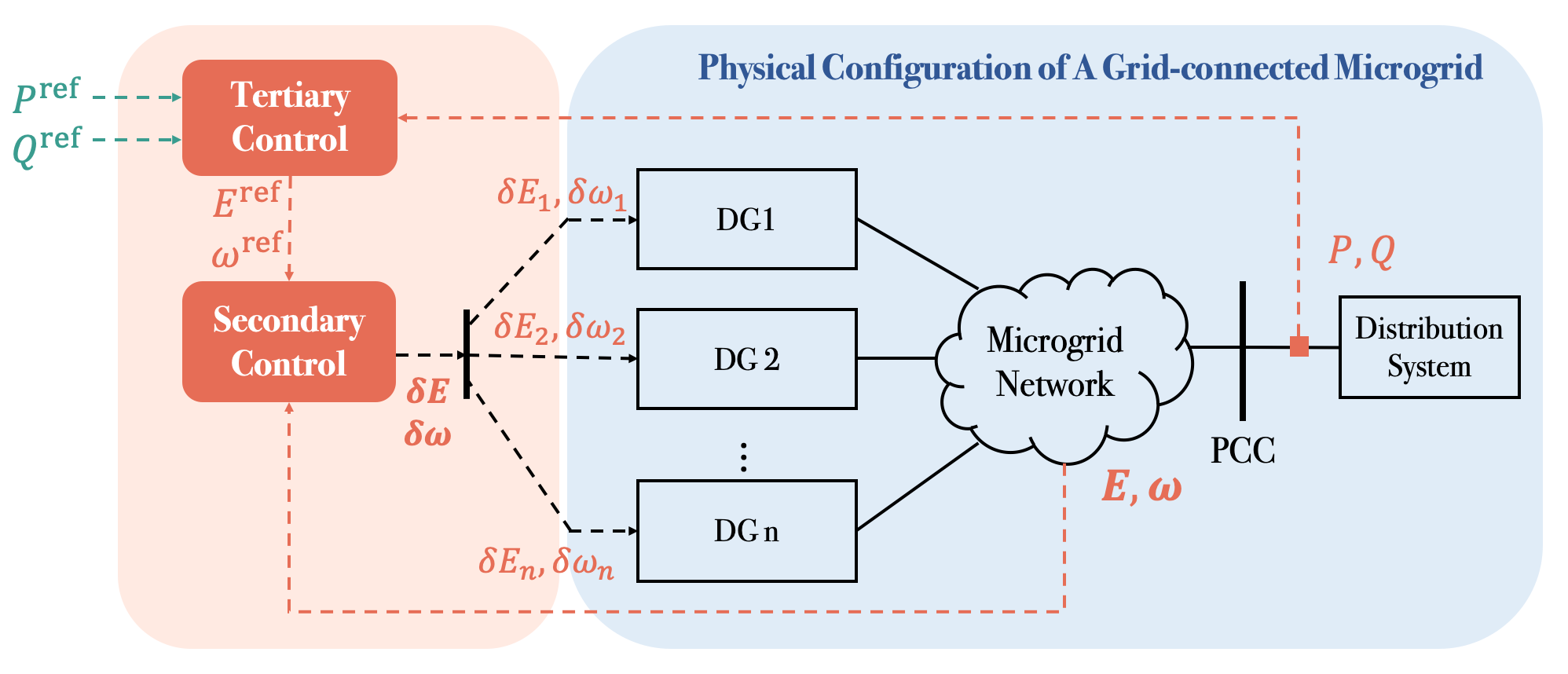}
    \caption{The secondary and tertiary control of an AC microgrid (Modified based on Fig. 5 of reference \cite{9528341} \copyright IEEE 2021)}
    \label{fig:secondary_control}
\end{figure}

The DERs, load, and energy storage within a microgrid are managed hierarchically. Taking an AC microgrid as an example, one may consider a hierarchical microgrid control scheme including device-level, secondary, and tertiary control. The device-level control of inverters, as in Section \ref{sec:IBRctrl}, 
can stabilize the frequency and voltage magnitudes in a decentralized manner.
In order to recover the frequency and voltage magnitudes to their nominal values, a secondary control is needed. Figure \ref{fig:secondary_control} shows a centralized secondary controller in a microgrid. The secondary controller aims to regulate the voltage magnitude and frequency at some critical microgrid nodes by tuning the setpoints of the grid-forming inverters. The setpoints of the secondary controller are given by the tertiary controller when the microgrid is in a grid-connected model, and they are determined by the microgrid itself when it is in an islanded mode. It is worth noting that the secondary controller can be implemented in a decentralized or distributed manner \cite{khayat2018decentralized, simpson2015secondary,dehkordi2016fully,dehkordi2016distributed,bolognani2013distributed}. The tertiary control regulates the real and reactive power flow between the microgrid and its host distribution system. The setpoints of the tertiary control are provided by the distribution system operators  \cite{SU201245,6872821}, and they are tracked by tuning the setpoints of the secondary controller. It should be noted that the \textcolor{black}{information and} control architecture is highly dependent on the communication infrastructure that allows for information exchange between each individual DER and the $\mu$MS, and between each $\mu$MS and the DISCO.


\textcolor{black}{To enhance the resilience and \textcolor{black}{economic} efficiency of microgrids, the microgrid control design entails aspects in two time scales: 1) how to design the equilibrium points in the steady-state time scale; and 2) how to achieve the pre-designed equilibrium in the fast time scale. The equilibrium point design concerns efficient DER allocation, and microgrid topology reconfiguration. The DER allocation problem determines the optimal sizing and placement of DERs for various purposes, such as minimizing operational costs under normal conditions while ensuring that critical loads are served during emergent conditions \cite{6377245,6208842,6894251,7559799,7879853,8019829}, and shortening restoration time \cite{7151852,8362717}. The microgrid topology reconfiguration problem determines how DERs are connected in order to enhance the distribution system's resilience \cite{7374730,7935520}.} 

Under disturbances, e.g., renewable fluctuations, and topology changes, the microgrid behaviors are stabilized around the pre-designed equilibrium points by the controllers embedded into the DERs and $\mu$MSes. References \cite{6200347,5546958,8892668} review the recent advances on \textcolor{black}{microgrid hierarchical control where generation resources at the grid edge are stabilized by lower-level controllers in a fast time scale, and higher-level controllers coordinate the generation resources in a slow time scale. The higher-level controllers can be implemented in a decentralized/distributed manner. A large body of literature (see \cite{9779512,8779692} and the references therein) addresses microgrid frequency and voltage regulation by introducing new distributed/decentralized control algorithms. Some research efforts handle the regulation problem by re-interpreting the grid dynamics in a new state space \cite{huang2022cyber, liu2013large,ilic1993simple,ILIC201852}} 
In order to make sure the controllers in a microgrid are well configured, it is critical for system planners and operators to establish the stability of a microgrid distribution system, under disturbances such as operational mode changes (e.g., one microgrid disconnects from its host grid), DERs' on/off status changes, and topology changes (e.g., line tripping). Simulation studies to investigate stability require small time steps of numerical integration due to the fast dynamics of IBRs, and so speeding up simulation is critical \cite{9387545,6923500,7365497}. A drawback of simulation-based methods is that they cannot rigorously establish stability, as only a finite set of disturbances are considered. On the other hand, analytical methods require a Lyapunov function to rigorously certify an equilibrium point's stability \cite{7419922,huang2021neural,9281628,7404044}, which is generally challenging \cite{FOUAD1988233, 481632, 9559389}. Recently, 
machine-learning techniques have been used to construct Lyapunov functions \cite{9559389}. 
The general limitation of existing analytical methods is that they typically use simplified models of inverter dynamics, e.g., \cite{9559389}, and
moreover require knowledge of the dynamics of all DERs in the microgrid, which
is not easily available.

While microgrids appear to be very promising, there is  limited experience about other services within microgrids, such as black-start without connecting to the main grid. Further, the above only discusses the stability of a microgrid-based distribution system. It is worth noting that the cybersecurity issues in the microgrid context are equally important, since the microgrid entails many control loops. Perhaps methods used for addressing the cybersecurity issues in transmission systems, e.g., the watermarking approach \cite{8345676}, can be migrated to distribution systems. 





\subsection{New Loops for Transportation Electrification}
Transportation electrification requires new control loops for the coordination of EV charging and discharging. Aggregators and EVs participate in this loop.
There is also the slower time-scale of planning required to determine EV charging station placement, in which DISCOs and EV charging stations participate. 

\subsubsection{Coordination of EV Charging and Discharging}


A large-scale deployment of EVs introduces both opportunities and challenges for distribution grid operation. On the one hand, the batteries of EVs can be used to provide grid services, such as optimizing load profiles \textcolor{black}{\cite{6313962}}, if they are well coordinated. On the other hand, 
simulation-based studies \cite{5356176, muratori2018impact} have shown that uncoordinated EV charging may jeopardize a distribution grid's security by incurring large line losses and large voltage deviations 
\cite{5356176}, and by decreasing the lifespan of grid equipment, such as transformers. 
In today's distribution systems, there is very limited coordination of EV charging and discharging. {\color{black}{A number of EV charging implementations are being researched and around the world some pilots have been conducted \cite{everoze2018v2g}. The most simple management approach perhaps involves the utility using a Time-of-Use (TOU) tariff specifically designed for EVs. This could encourage customers to move their charging from peak to off-peak hours \cite{mckinsey2018}. In this case the control approach is fairly simple since smart meters which can continuously monitor electricity consumption have already been rolled out in many developed countries, and the monthly bill can be adjusted based on the TOU tariff. Next is unidirectional controlled charging or V1G, where the rate of EV charging is adjusted based on control signals from the system operator. The system operator would send the control signal either directly to the software in the EV (where enabled) or to an aggregator, and then would need aggregation-level telemetry from the aggregator in order to get visibility. The aggregator is responsible to the system operator for providing accurate response (within requirements) and also ensuring the privacy of individual EV owners.

 Vehicle-to-home (V2H) or vehicle-to-building (V2B) options are also being considered. For instance, with V2H technology EV owners can utilize their vehicles as a source of backup power in case of grid disruptions \cite{Turker2018}. While the above methods involve a one-way flow of electricity, bi-directional V2G technology allows the energy stored in EV batteries to be utilized for many grid supporting services \cite{Siddhartha2013, Amamra2019, Sami2019}. V2G technology could help in renewable energy integration by providing ancillary services \cite{Iqbal2018}. EVs could also help provide peak demand management and voltage support for the distribution grid \cite{Kesler2014}. The control mechanism for V2G is perhaps the most complex since it requires understanding/predicting the customer charging schedules in order to deliver reliable grid services. Further, there are limits on how many EVs the grid can charge at the same time without requiring expensive distribution system upgrades. Customers would also need to be given options (say through a smart phone app) to be able to set their desired end State of Charge (SOC) for a charging session.}}
\textcolor{black}{For the future grid, it can be envisioned that EV chargers will have the capability to communicate with an aggregator. Further, there may exist contractual agreements between EV owners and the aggregator that allow the aggregator to control the charging rates of the EV chargers within a scheduling horizon, e.g., 24 hours. The aggregator would take into account the requirements on availability and SOC specified by individual EV owners \cite{garcia2014plug,ma2016efficient,6313962}. }
The aggregator may seek to minimize the line loss of the distribution system, to fill the valley of a load profile, or to establish a tradeoff between costs of energy and battery degradation 
\cite{5356176, 6313962, ma2016efficient, garcia2014plug, 8967190}.



{\color{black}{There are still a number of unanswered questions. Who is the best entity to provide the control signal - ISO, DISCO, or third party aggregator? There are also regulatory barriers for EVs to participate in many electricity markets.}}



\subsubsection{EV Charging Station Placement}

The adoption of EVs will need widespread charging infrastructure to be developed, at different charging levels, and for a variety of transportation needs ranging from local commutes to inter-city travel to commercial transportation. 
Sufficient deployment of charging infrastructure is a prerequisite for large-scale deployment of EVs.
A natural question is how to efficiently deploy EV charging stations in a distribution power system. This problem is termed as the EV charging station placement problem which aims to determine the locations and capacities of the EV charging stations \textcolor{black}{\cite{6879337}}. 

Currently there is little coordination of the electric and transportation infrastructures for EV charging station (EVCS) placement. This is a major area of research opportunities for the future grids with massive numbers of EVs. For the EVCS placement problem, EVCS planners aim to determine the sites and capacity of EVCSs. The problem can be formulated as an optimization problem with an objective function that \textcolor{black}{reflects the costs including EVCS construction, distribution system expansion, and voltage violation.} 
The constraints are related to EV charging demand in the area under study, including power distribution network limits, and budget limits. A large body of literature is devoted to coordinating the electric and transportation infrastructure by considering various practical factors
 \cite{6362255,6710196,8673613,sun2020integrated,kchaou2021charging}.

{\color{black}{The planning for EV charging infrastructure may have to be done at multiple levels. At the regional level, to support fast charging stations on highways and fleet charging for trucks there would have to be coordination between electricity transmission and transportation infrastructure planning. At the local level, distribution system upgrades would have to done considering the increasing adoption of EVs as well as the potential for two-way power flow to enable V2G applications.}}

\subsection{Demand Response}

The objective of conventional power grid operation is to make generation track load, since fossil-fuel generators are dispatchable, while loads are traditionally not considered as finely controllable. In the future grid, large-scale integration of renewables would render the generation side less dispatchable, but intelligence at the grid edges, such as smart meters and Internet of Things (IoTs), enhance load controllability \cite{5643088}. This enables the load to track the generation. Demand response programs make it possible to unlock the flexibility at the load side in order to maintain load-generation balance. Demand response (DR) programs are designed for changing customers' electricity usage patterns\cite{goldman2010coordination}. DR programs can be broadly categorized into 1) price response and 2) direct load control \cite{5643088}. The price response approach changes the electricity usage patterns by using price signals or other incentives. Examples of price response include market-index retail plans designed by utilities, time-of-use \cite{borenstein2002dynamic}, critical peak pricing \cite{borenstein2002dynamic}, and coupon-based programs \cite{ming2020prediction}. To illustrate the basic mechanism of price response, \textcolor{black}{the coupon-based program \cite{ming2020prediction} is taken as an example.} As an aggregator purchases electricity from the wholesale market with a time-varying price, but typically sells electricity to customers at a fixed rate, the aggregator's goal is to reduce the electricity consumption whenever the wholesale price is high. 
To this end, one possible mechanism is where the aggregator can leverage a coupon-based demand response program  \cite{ming2020prediction}. The aggregator first predicts \textcolor{black}{the} electricity price several hours ahead, say, 2-hours ahead, and checks whether the price forecast exceeds a threshold. If yes, the aggregator can inform its customers via mobile phone applications that they will obtain a certain number of lottery tickets if they can decrease \textcolor{black}{their} electricity usage by a certain amount within a given period of time. The number of lottery tickets issued to a customer can be based on the price forecast and energy saving behavior of the customer \cite{ming2020prediction}.

\begin{figure}
    \centering
    \includegraphics[width = \linewidth]{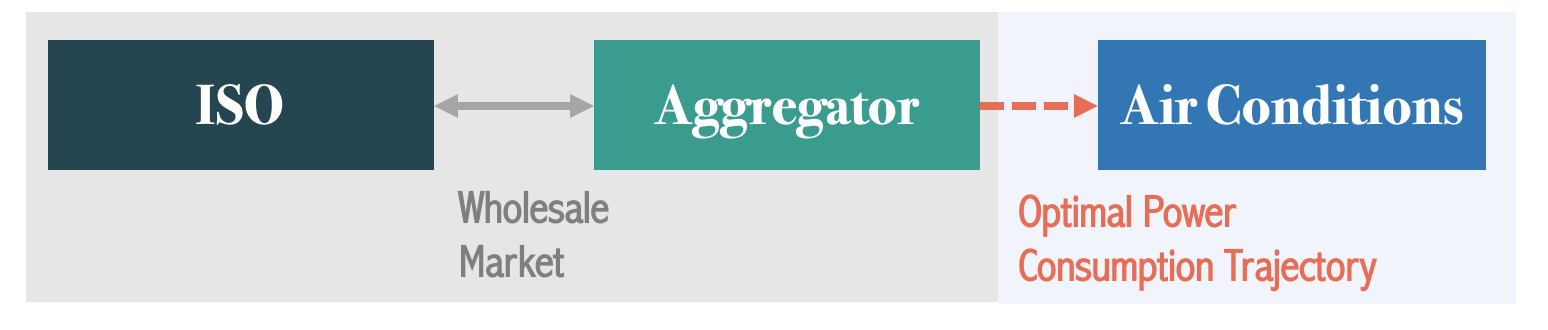}
    \caption{A two-layer thermal inertial load control system}
    \label{fig:LSE}
\end{figure}

Direct load control requires aggregators (or Load Serving Entities, LSEs) to change the setpoints of customers' equipment, e.g., air conditioners \cite{7742416}, refrigerators, and swimming pool pumps \cite{modarresi2018reserves}. As an example, reference \cite{7742416} proposes a two-layer direct load control strategy which is shown in Figure \ref{fig:LSE}. Suppose that the aggregator can control the setpoints of many air conditioners. In the first layer, given a planning horizon, an aggregator participates in the wholesale electricity market and determines an optimal power consumption trajectory over the planning horizon with an objective of minimizing costs from purchasing energy in the wholesale market. Once the optimal power consumption trajectory is determined, the second layer can change the setpoint of each AC in order to drive the total power consumption to track the optimal power consumption trajectory, while making sure that the comfort temperature for each individual home is maintained.

The adoption of both direct load control and time-varying rates is currently in a nascent stage despite many pilot programs tested by a number of utilities. As grid-edge technologies proliferate the potential for such programs could increase significantly.

\subsection{New Business Model for Utilities with Prosumers}
The societal choice of decarbonization and electrification will lead to the growth of DERs. Hence many traditional consumers could become \textit{prosumers} who produce their own energy, store, and sometimes sell it to the grid. The rise in DER penetration poses a threat to the revenue streams of utilities. This is an important issue since utilities need to be compensated for the investments they have made in the distribution system infrastructure as well as the operating costs they incur. The current rate structure depends heavily on volumetric charges. However, by installing solar panels and storage devices, customers may be able to significantly reduce their consumption from the grid. In the short term utilities might be able to compensate the loss of revenue by increasing the fixed charges. However, this brings with it the risk of ``grid defection" by customers and a resulting downward spiral of the utility revenue stream.

There is a need to adapt the rate designs in order to compensate a utility for the network it provides, as well as other services, for instance maintaining power quality. Distribution utility business models will need to evolve in order to handle large numbers of DER prosumers \cite{venkatraman2021smart}.

Historically the direction of power flow was uni-directional -- from the sub-station to the customer. However, with DERs, this is likely to change and will require the power flow, communications and control signals to all be bidirectional. It is important to establish a level playing field
between traditional generators, network providers and
DER aggregators for the provision electricity services \cite{MIT_UF}.

{\color{black}{
\subsection{Multi Energy Systems}
Multi Energy Systems (MES) that consider interactions between various energy vectors (electricity, heating/cooling and gas) are gaining attention \cite{Mancarella2014}. The concept of MES can be extended to multiple spatial levels ranging across buildings, districts, cities and regions. For instance, building heating and cooling technologies such as Combined Heat and Power (CHP), heat pumps and air-conditioning have interactions. The emergence of EVs is leading to coupling between regional electricity and transportation networks. The joint modeling and optimization of multi-energy systems is a challenging problem, but one that could improve energy efficiency and reduce emissions. Such systems can be optimized to provide energy and balancing services in electricity markets \cite{Defossez2018}. Stochastic optimization approaches have been used to optimize the operation and design of MES \cite{Najafi2016, Pazouki2016}. Also robust optimization has been applied for the optimal selection, sizing, and operation of MES \cite{Gabrielli2019}. Economic Model Predictive Control can be used to control MES under uncertain weather, loads and renewable power in order to minimize costs \cite{Blaud2020}. A layered \textcolor{black}{information and} control architecture for the control of a multi energy system is presented in \cite{Li2022}.
}}
%



\section{Concluding Remarks}\label{Sec:Conclusion}
\textcolor{black}{Finally, the issue of {\textit{extensibility}} is discussed here. Extensibility is} an important property of a design that allows the future incorporation of new capabilities or functionalities. The \textcolor{black}{information and} control architecture of the \textcolor{black}{current power grid} was designed to have the important properties of a reliable and efficient system, such as fault tolerance, scalability, \textcolor{black}{and} stability.
The question before us today is whether the original design is also extensible to allow for large penetration of renewables along with other compensatory strategies on the demand and storage side.
This paper
provides constructive suggestions towards achieving this goal at several levels, and thereby
argues that the current design is indeed extensible.
Specifically, \textcolor{black}{several new control loops have been articulated} for the future electric energy systems aimed at achieving the new societal goals of decarbonization and electrification of transportation.

Some new research challenges are described as follows from the viewpoint of \textcolor{black}{system} architecture design. The first challenge is how to balance the pragmatic view of a system with large sunk investment \textcolor{black}{versus} a greenfield view of the system. The architecture of the electric energy system has evolved over the past century with many loops of control, communication, and computing technologies intertwined with the physical system. Thus the challenge is how to design new control loops that are backward compatible with existing loops.

The second challenge is how to provide a coherent viewpoint of this complex system across multiple temporal and spatial scales, with some formal verification or some form of guarantee of performance. The assumptions of time-scale and spatial-scale separation will need to be mathematically re-visited with the increasing level of complexity involved with this system, particularly in light of new distributed energy technologies and intelligence at the grid-edge. 

A third challenge is how to translate these
control loop innovations into practical implementation. Many of the innovations of architectural design will need to be blended with rapid prototyping, trials and errors, and feedback design. Such intermediate efforts between research and implementation will need to be facilitated to more effectively and nimbly adapt to the changing nature of the complex energy systems.

The fourth challenge is how to provide architectural design principles \textcolor{black}{for} dynamically changing societal objective\textcolor{black}{s}. As of now the pivotal concerns of the electric energy system are how to decarbonize \textcolor{black}{and} how to provide resilient electricity services. However, societal needs and objectives may further change over a long course of time. Architecture design will need to be re-visited and adapted accordingly at a longer time scale trajectory. 

\printnomenclature

\IEEEpeerreviewmaketitle{}

\bibliographystyle{IEEEtran}
\bibliography{main.bib}

\begin{IEEEbiography}
[{\includegraphics[width=1in,height=1.25in,clip,keepaspectratio]{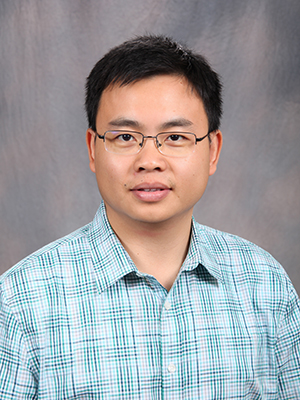}}] 
{Le Xie} (Fellow, IEEE) received his B.E. degree in Electrical Engineering from Tsinghua University, Beijing, China, in 2004, an M.S. degree in Engineering Sciences from Harvard University,
Cambridge, MA, USA, in 2005, and a Ph.D. degree from Carnegie Mellon University, Pittsburgh, PA,
USA, in 2009. He is currently a professor with the Department of Electrical and Computer Engineering, Texas A\&M University, College Station, TX, USA. His research interests include modeling and control of large-scale complex systems, smart grids application with renewable energy resources, and electricity markets.
\end{IEEEbiography}

\vskip -2\baselineskip plus -1fil

\begin{IEEEbiography}
[{\includegraphics[width=1in,height=1.25in,clip,keepaspectratio]{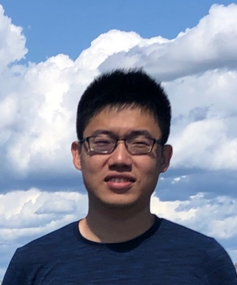}}] 
{Tong Huang} (Member, IEEE) is an Assistant Professor in the Department of Electrical and Computer
Engineering at San Diego State University (SDSU). Before joining SDSU, he was a postdoctoral
associate at MIT. He received his Ph.D. degree from Texas A\&M University in 2021. His
industry experience includes an internship with ISO-New England in 2018 and an internship
with Mitsubishi Electric Research Laboratories in 2019. As the first author, he received two Best
Paper Awards at the 2020 IEEE PES General Meeting and the 54-th Hawaii International
Conference on System Sciences. His research focuses on data analytics, cyber security, and
modeling and control of power grids with deep renewables.
\end{IEEEbiography}

\vskip -2\baselineskip plus -1fil
\begin{IEEEbiography}[{\includegraphics[width=1in,height=1.25in,clip,keepaspectratio]{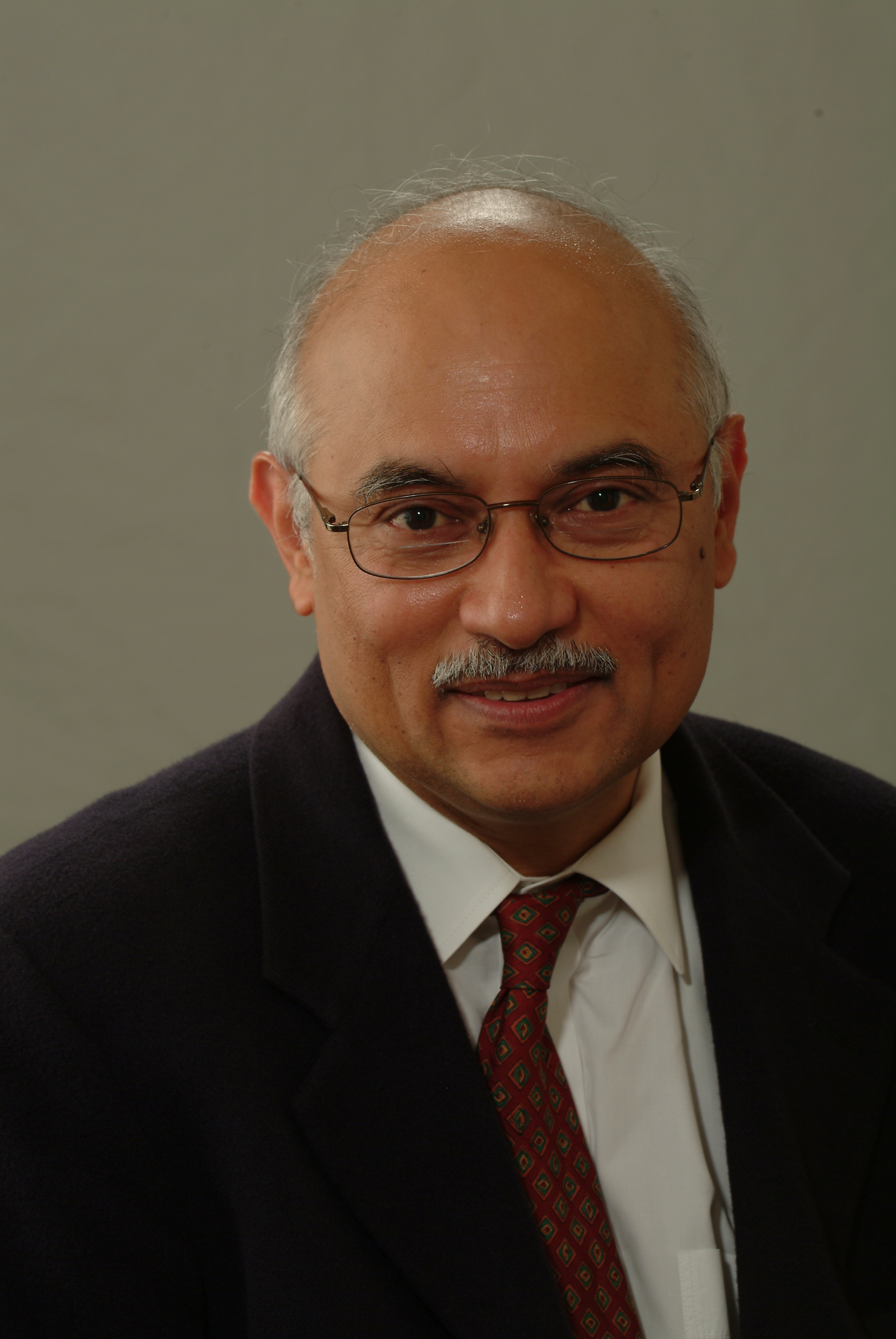}}]{P. R. Kumar} (Life Fellow, IEEE) B. Tech. (IIT Madras, 73), D.Sc.(Washington University, St. Louis, 77), was a faculty member at UMBC (1977-84) and Univ. of Illinois, Urbana-Champaign (1985-2011). He is currently at Texas A\&M University. He is a member of the US National Academy of Engineering, The World Academy of Sciences, and the Indian National Academy of Engineering. He was awarded a Doctor Honoris Causa by ETH, Zurich. He has received the Alexander Graham Bell
Medal of IEEE, the IEEE Field Award for Control Systems, the Donald P. Eckman
Award of the AACC, the Fred W. Ellersick Prize of the IEEE Communications Society, the Outstanding Contribution Award of ACM SIGMOBILE, the INFOCOM Achievement Award, the SIGMOBILE Test-of-Time Paper Award, and the Outstanding Contribution Award of COMSNETS. He is a Fellow of IEEE and ACM Fellow. He was Leader of the Guest Chair Professor Group on Wireless Communication and Networking at Tsinghua University, and a D. J. Gandhi Distinguished Visiting Professor at IIT Bombay. He is
an Honorary Professor at IIT Hyderabad. He was awarded the Distinguished Alumnus Award from IIT Madras, the Alumni Achievement Award from Washington Univ., and the Daniel Drucker Eminent Faculty Award from the College of Engineering at the Univ. of Illinois. His current research is focused on machine/reinforcement learning, power systems, wireless networks, security of cyber-physical systems, privacy, unmanned aerial vehicle traffic management, and millimeter wave 5G.
\end{IEEEbiography}
\vskip -2\baselineskip plus -1fil
\begin{IEEEbiography}[{\includegraphics[width=1in,height=1.25in,clip,keepaspectratio]{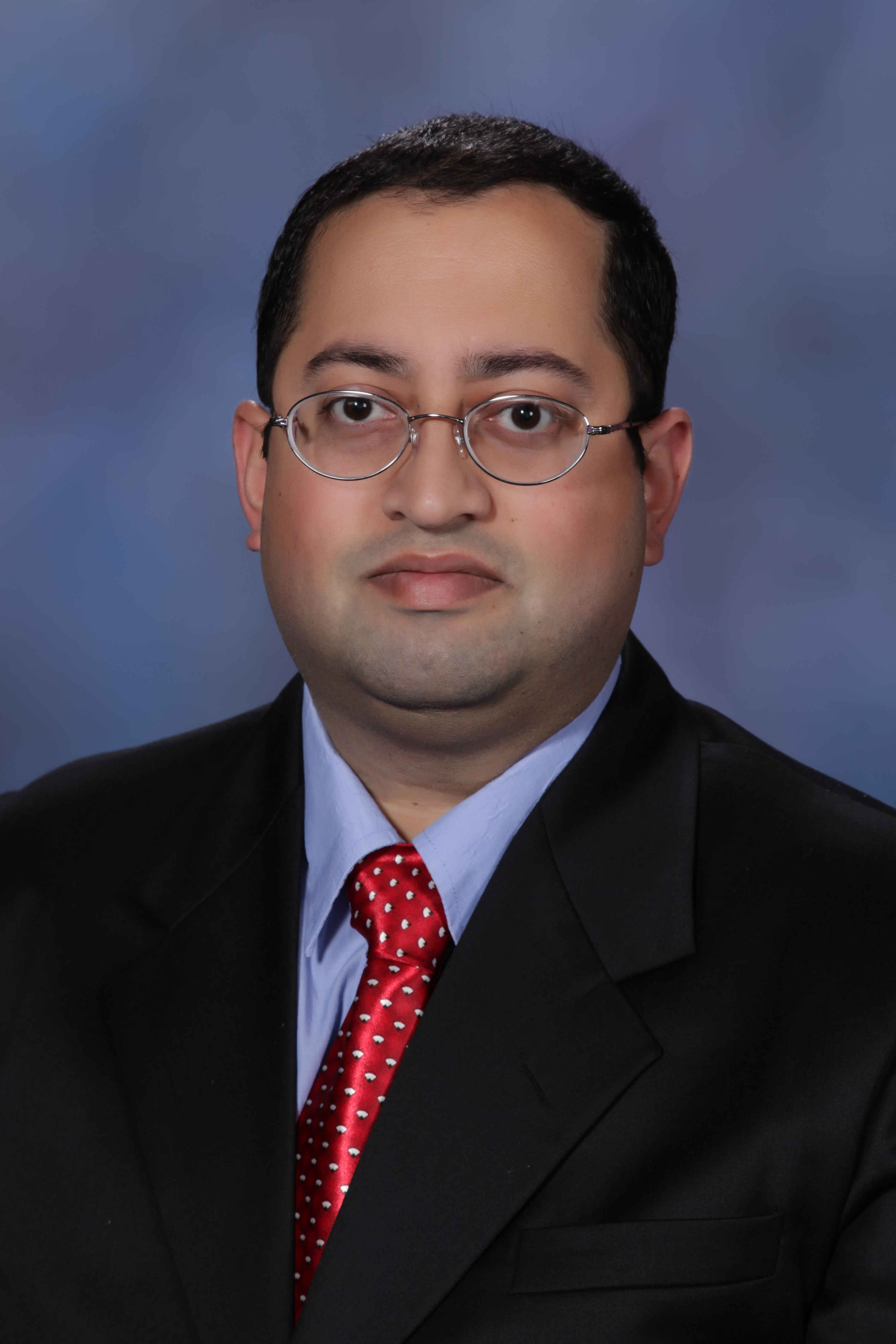}}]{Anupam A. Thatte} (Member, IEEE) received the B.E. degree in electrical engineering from Pune University, Pune, India, the M.S. degree in electrical and computer engineering from Carnegie Mellon University, Pittsburgh, PA, USA, and the Ph.D. degree in electrical and computer engineering from Texas A\&M University, College Station, TX, USA.
He is currently working for the Midcontinent Independent System Operator, Carmel, IN, USA. His research interests include modeling and control of power systems, grid integration of renewable energy and electricity markets.
\end{IEEEbiography}

\vskip -2\baselineskip plus -1fil

\begin{IEEEbiography}[{\includegraphics[width=1in,height=1.25in,clip,keepaspectratio]{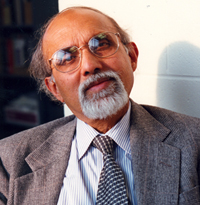}}]{Sanjoy K. Mitter} (Life Fellow, IEEE) received his Ph.D. degree from the Imperial College of Science and Technology in 1965. He taught at Case Western Reserve University from 1965 to 1969. He joined MIT in 1969 where he has been a Professor of Electrical Engineering since 1973. He was the Director of the MIT Laboratory for Information and Decision Systems from 1981 to 1999. He has also been a Professor of Mathematics at the Scuola Normale, Pisa, Italy from 1986 to 1996. He has held visiting positions at Imperial College, London; University of Groningen, Holland; INRIA, France; Tata Institute of Fundamental Research, India and ETH, Zürich, Switzerland; and several American universities. Professor Mitter is the recipient of the IEEE Eric E. Sumner Award for 2015. Also in 2015, Professor Mitter was elected a Foreign Fellow of the Indian National Academy of Engineering. He was the Ulam Scholar at Los Alamos National Laboratories in April 2012 and the John von Neumann Visiting Professor in Mathematics at the Technical University of Munich, Germany from May-June 2012. He was awarded the AACC Richard E. Bellman Control Heritage Award for 2007. He was the McKay Professor at the University of California, Berkeley in March 2000, and held the Russell- Severance-Springer Chair in Fall 2003. He is a Fellow of the IEEE and IFAC. He is the winner of the 2000 IEEE Control Systems Award. He was elected a Foreign Member of Istituto Veneto di Scienze, Lettere ed Arti in 2003. In 1988, he was elected to the National Academy of Engineering.
\end{IEEEbiography}


\end{document}